\documentclass[12pt]{article}
\usepackage[english]{babel}
\usepackage[latin1]{inputenc}
\usepackage{amsfonts,amssymb,amsmath, epsfig}
\usepackage{color,graphicx,graphics,psfrag}
\usepackage{amsmath,amstext,amssymb,amsfonts, amscd}
\usepackage{hyperref}           

\def\Box{\leavevmode\vbox{\hrule
     \hbox{\vrule\kern4pt\vbox{\kern4pt}%
           \vrule}\hrule}}
\def\blackbox{\leavevmode\vrule height 5pt width 4pt depth 0pt\relax}
\def\endproof{\null\hfill {$\blackbox$}\bigskip}

\newcounter{appendix}
\def\appendix{\advance\c@appendix by 1
   \def\thesection{\Alph{section}}
   \ifnum\c@appendix=1 \setcounter{section}{-1} \fi
   \@startsection {section}{1}{\z@}{-3.5ex plus -1ex minus 
   -.2ex}{2.3ex plus .2ex}{\Large\bf}}

\def\paragraph#1{{\bf #1\ }}

\newtheorem{lemma}{Lemma}[section]

\newtheorem{definition}[lemma]{Definition}

\newtheorem{remark}{Remark}[section]

\title{A hierarchy of heuristic-based models of crowd dynamics} 
\author{P. Degond$^{1,2}$, C. Appert-Rolland$^{3,4}$, M. Moussaid$^{5}$, \\J. Pettre$^{6}$, G. Theraulaz$^{7,8}$} 
\date{} 
\begin{document}

\maketitle

\vspace{0.5 cm}

\begin{center}
1-Université de Toulouse; UPS, INSA, UT1, UTM ;\\ 
Institut de Mathématiques de Toulouse ; \\
F-31062 Toulouse, France. \\
2-CNRS; Institut de Mathématiques de Toulouse UMR 5219 ;\\ 
F-31062 Toulouse, France.\\
email: pierre.degond@math.univ-toulouse.fr
\end{center}

\begin{center}
3- Laboratoire de Physique Théorique, Université Paris Sud, \\
bâtiment 210, 91405 Orsay cedex, France\\
4- CNRS, UMR 8627, Laboratoire de physique théorique, 91405 Orsay, France\\
email: Cecile.Appert-Rolland@th.u-psud.fr
\end{center}

\begin{center}
5-1 Center for Adaptive Behavior and Cognition, Max Planck Institute for Human Development, Lentzeallee 94, 14195 Berlin, Germany \\
email: moussaid@mpib-berlin.mpg.de
\end{center}

\begin{center}
6-INRIA Rennes - Bretagne Atlantique, Campus de Beaulieu, 35042 Rennes, France \\
email: julien.pettre@irisa.fr
\end{center}

\begin{center}
4-Centre de Recherches sur la Cognition Animale, UMR-CNRS 5169, \\ Université Paul Sabatier, Bât 4R3, \\
118 Route de Narbonne, 31062 Toulouse cedex 9, France. \\
8- CNRS, Centre de Recherches sur la Cognition Animale, F-31062 Toulouse, France \\
email: theraula@cict.fr
\end{center}

\vspace{0.5 cm}
\begin{abstract}
We derive a hierarchy of kinetic and macroscopic models from a noisy variant of the heuristic behavioral Individual-Based Model of \cite{Moussaid_etal_PNAS11} where the pedestrians are supposed to have constant speeds. This IBM supposes that the pedestrians seek the best compromise between navigation towards their target and collisions avoidance. We first propose a kinetic model for the probability distribution function of the pedestrians. Then, we derive fluid models and propose three different closure relations. The first two closures assume that the velocity distribution functions are either a Dirac delta or a von Mises-Fisher distribution respectively. The third closure results from a hydrodynamic limit associated to a Local Thermodynamical Equilibrium. We develop an analogy between this equilibrium and Nash equilibia in a game theoretic framework. In each case, we discuss the features of the models and their suitability for practical use. 
\end{abstract}

\medskip
\noindent
{\bf Acknowledgments:} This work has been supported by the French 'Agence Nationale pour la Recherche (ANR)' in the frame of the contracts 'Pedigree' (ANR-08-SYSC-015-01) and 'CBDif-Fr' (ANR-08-BLAN-0333-01)


\vskip 0.4cm

\setcounter{equation}{0}
\section{Introduction}
\label{sec_intro}

Understanding and predicting crowd behavior is an extremely important issue in our societies. Public safety concerns have raised considerably after recent major crowd disasters \cite{Ngai_etal_DisasterMed09}. Public authorities are challenged by increasingly large crowds attending mass events such as sports gatherings. Economic stakes related to crowd management are equally high, as increasing the efficiency of pedestrian infrastructures have important returns in terms of business. 

To achieve a better comprehension of crowd behavior and increase the reliability of predictions, numerical modeling and simulation is playing an ever-growing role. 

A recent review on crowd modeling can be found in \cite{Bellomo_Dogbe_SIAMRev11}. The most widely used crowd simulation models are Individual-Based Models (or IBM), such as Rule-Based models \cite{Reynolds_ProcGameDev99}, mechanical models \cite{Helbing_BehavSci91, Helbing_Molnar_PRE95, Helbing_Molnar_SelfOrganization97}, traffic following models \cite{Lemercier_etal_Eurographics12}, optimal control theory models, \cite{Hoogendoorn_Bovy_OptControlApplMeth03}, Cellular-Automata  \cite{Nishinari_etal_IEICETranspInfSyst04} and Vision-Based models \cite{Guy_etal_Siggraph09, Huang_etal_RoboticsAutonomSyst06, Ondrej_etal_Siggraph10, Paris_etal_Eurographics07, Pettre_etal_Siggraph09, Vandenberg_Overmars_IntJRoboticsRes08}. The present paper relies on \cite{Moussaid_etal_PNAS11} which is detailed below. Continuum models (CM) are based on a fluid dynamics approach \cite{Helbing_ComplexSyst92, Henderson_TranspRes74}. Other approaches through optimal control theory \cite{Huang_etal_TranspResB09, Hughes_TranspResB02, Hughes_AnnRevFluidMech03, Jiang_etal_PhysicaA10} or exploiting the analogy with car traffic \cite{Appert-Rolland_etal_NHM11, Bellomo_Dogbe_M3AS08, Berres_etal_NHM11, Colombo_Rosini_MMAS05, Coscia_Canavesio_M3AS08, Motsch_etal_friction, Piccoli_Tosin_ContMechThermo09} have also been developed. For dense crowds, the handling of the volume exclusion constraint has led to several specific works \cite{Degond_Hua_arXiv:1207.3522, Degond_etal_JCP11, Maury_etal_NHM11}. Existence theory for some CM of crowds can be found in \cite{DiFrancesco_etal_JDE11}. 

Kinetic Models (KM) are intermediate between IBM and CM. As pointed out in the review \cite{Bellomo_Dogbe_SIAMRev11}, there are quite few KM of crowds (see an example in \cite{Bellomo_Bellouquid_MathModelCollectivBehav10}). IBM, KM and CM constitute a hierarchy of models in the sense that each level can be deduced from the previous one by a model reduction methodology. Indeed, KM deal with the one-particle probability distributions of IBM. Such a description ignores correlations between the particles (which are described by joint probability distributions of $k$ particles for $k \geq 2$) and is therefore a reduced description of the IBM. CM are deduced from KM by taking averages over the velocity variable (see section \ref{subsubsec_monokinetic_derivation}). Therefore, CM involve a reduced description of the velocity statistics of the KM to a small number of its moments.

In general, CM or KM are more efficient than IBM for large crowds because their computational time does not increase with the number of agents. However, they suffer from different drawbacks, such as a reduced validity range due to the necessary recourse to closure relations, as detailed below. Nonetheless, CM are invaluable tools for large-scale analysis and prediction of crowd behavior. Therefore, it is important to firmly base the derivation of CM on their small-scale IBM counterpart. The literature on the derivation of CM from microscopic models (IBM or CA) is scarce (see e.g. \cite{AlNasur_Kachroo_IEEEITSC06, Burger_etal_KRM11, Chertock_etal_preprint12, Cristiani_etal_MMS11}). The present paper addresses this question and intends to propose a hierarchy of KM and CM based on the IBM developed in \cite{Moussaid_etal_PNAS11}. 

The psychological literature shows that pedestrians can estimate the positions and velocities of moving obstacles such as other pedestrians with fairly good accuracy \cite{Cutting_etal_PsychRev95}. Therefore, the subjects  are able to process this information in order to determine the dangerousness level of an encounter \cite{Warren_Fajen_OpticFlow04}. Taking these considerations into account, the heuristic-based model of \cite{Moussaid_etal_PNAS11} proposes that pedestrians follow a heuristic rule composed of two phases: a perception phase and a decision-making phase. In the perception phase, the subjects make an assessment of the dangerousness of the possible encounters in all the possible directions of motion. In the decision-making phase, they turn towards the direction which maximizes the distance walked towards their target while avoiding encounters with other pedestrians. In this sense, in the game theoretical sense, the pedestrians choose the Nash equilibrium as the new direction of motion. Game theoretical approaches of traffic have already been considered (see e.g. \cite{Rosenthal_IntJGameTheory73}). 

The goal of this paper is to derive a CM from this heuristic-based IBM. With this aim, the time-discrete IBM of \cite{Moussaid_etal_PNAS11} is first replaced by a time-continuous IBM
and noise is added to account for some uncertainty in the pedestrian velocity. 
From this time-continuous IBM, a KM is introduced. The KM describes the evolution of the probability distribution function of pedestrians in a phase space composed of position, velocity and target direction. For the sake of simplicity, we assume that the pedestrian speed remains constant and we discard any slowing down induced by close encounters. We do not develop any rigorous theory of the passage from the IBM to the KM \cite{Bolley_etal_AML12, 
Carlen_etal_arXiv:1109.4538}. 

The passage from the KM to the CM is realized by taking the velocity moments of the distribution function. In doing so, some closure relations are needed otherwise the hierarchy of moments is infinite. We propose three distinct closure relations. The first one assumes a monokinetic distribution function. In other words, the velocity distribution is assumed to be a Dirac delta at the mean velocity. Such a monokinetic Ansatz can only be valid in the noiseless case but provides an exact solution of the KM. In the second closure relation, the velocity distribution function is supposed to be a von Mises-Fisher (VMF) distribution
\cite{Watson_JAP82}. It is adapted to situations where the noise is non-zero. 
In these first two closures, the resulting macroscopic model is a system consisting of a mass conservation equation and an evolution equation for the mean velocity of each ensemble of pedestrians sharing the same target direction.

The last closure is based on a formal hydrodynamic limit. It can be performed in the restrictive situation where (i) the pedestrian interactions can be approximated by spatially local ones and (ii) the interaction region of the subjects is isotropic (i.e. there is no blind zone behind the subjects). The closure relies on a Local Thermodynamical Equilibrium (LTE)
obtained through the solution of a fixed point equation. It expresses that each ensemble of pedestrians has found its optimal mean velocity in the midst of the other ones, i.e. is a Nash equilibrium in the game-theoretical sense. This example fits in the general framework relating kinetic theory and game theory which can be found in \cite{Degond_etal_arXiv:1212.6130} and which bears analogies with the theory of Mean-Field Games \cite{Lasry_Lions_JapanJMath07}. 
In a companion paper \cite{Degond_etal_synthetic}, the same methodologies are applied to the model of \cite{Ondrej_etal_Siggraph10} based on a mechanical view of pedestrian encounters. 

The outline of the paper is as follows. In section \ref{sec:heuristic} we review the IBM of \cite{Moussaid_etal_PNAS11}. In sections \ref{sec:kinetic} and \ref{sec:macro}, we successively derive the KM and the CM with the three possible closure relations. The resulting models are discussed in \ref{sec:discussion}. 
Finally, a conclusion is drawn in section \ref{sec:conclu}.

\setcounter{equation}{0}
\section{The Heuristic-Based model of pedestrian dynamics}
\label{sec:heuristic}

\subsection{Principles}
\label{subsec:heuristic_principles}

The heuristic-based model of \cite{Moussaid_etal_PNAS11} proposes that pedestrians follow a  rule composed of two phases: a perception phase and a decision-making phase. 

\begin{figure}
\begin{center}
\input{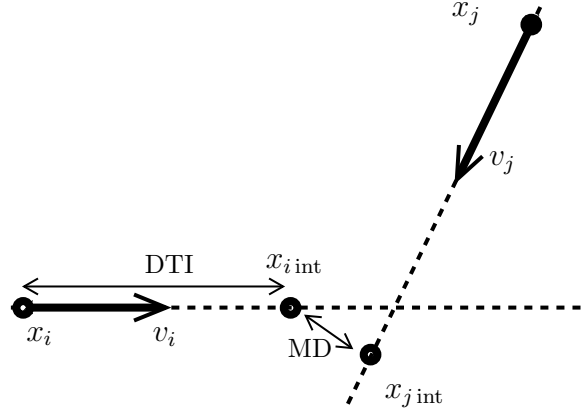}
\caption{The Minimal Distance MD and the Distance-To-Interaction (DTI). The MD is the smallest distance which separates the two pedestrians $i$ and $j$ supposing that they cruise on a straight line at constant velocities $v_i$ and $v_j$. The point on pedestrian $i$'s trajectory where the minimal distance is attained is the interaction point $x_{i \, \mbox{\scriptsize{int}}}$ of pedestrian $i$ in his interaction with pedestrian $j$. The MD is the distance between $x_{i \, \mbox{\scriptsize{int}}}$ and $x_{j \,  \mbox{\scriptsize{int}}}$. The DTI is the distance which separates the current pedestrian position $x_i$ to the interaction point $x_{i \, \mbox{\scriptsize{int}}}$. The Time-To-Interaction (TTI) is the time needed by pedestrian $i$ to reach the interaction point from his current position. Clearly, TTI $=$ DTI$/|v_i|$. }
\label{Fig_collision_1_2}
\end{center}
\end{figure}

In the perception phase, the key observables are the distance-to-interaction (DTI), the time-to-interaction (TTI) and the minimal distance (MD). Let us first examine a binary encounter with another pedestrian (see Fig. \ref{Fig_collision_1_2}). 

\begin{definition}
Consider a pedestrian (the subject) at a given location and time moving with a given velocity. Suppose that this pedestrian interacts with a single other pedestrian (the collision partner) who possibly has a different velocity but whose location at the same time is close. In this encounter, we define the following quantities:

\noindent
(i) The interaction point is the point on the subject's future or past trajectory where the distance to the collision partner is minimal, assuming that both agents move in straight line with a constant speed. 

\noindent
(ii) The Minimal Distance (MD) is this minimal distance between the subject and his collision partner. 

\noindent
(iii) The Distance To Interaction (DTI) is the distance which separates the subject's current position to the interaction point. The DTI is counted positive if the interaction point will be reached in the future of the subject and negative if the interaction point has been crossed in the past. 

\noindent
(iv) The Time-To-Interaction (TTI) is the time needed by the subject to reach this interaction point from his current position (counted positive if this time belongs to the future of the subject and negative if it belongs to the past). 

\label{def_bin_interaction}
\end{definition}

\begin{remark}
(i) If the TTI is negative (i.e. the interaction point has been reached in the past and the subject and his collision partner are now moving away from each other), or if the MD is above a certain threshold (equal to the subjects' diameter, possibly augmented by some safe-keeping distance), then, no interaction occurs and the DTI is set to infinity. 

\noindent
(ii) Because the subjects have supposedly perfect knowledge of their own and partner's positions and velocities, we assume that they are able to estimate the DTI, TTI and MD with perfect accuracy. 

\label{rem_bin_interaction}
\end{remark}

\begin{figure}
\begin{center}
\input{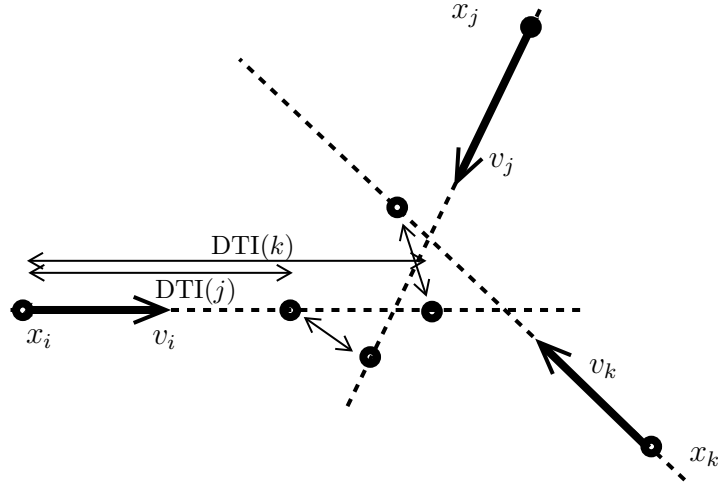}
\caption{Perception phase: the Distance-To-Interaction (DTI) of a given pedestrian in the case of several simultaneous encounters is the minimum of the DTI of the individual encounters. In this figure, the DTI of pedestrian $i$ with pedestrian $j$ (denoted by DTI$(j)$) is smaller than that with pedestrian $k$ (denoted by DTI$(k)$). Therefore, the DTI of pedestrian $i$ is DTI$(j)$. }
\label{Fig_collision_1_3}
\end{center}
\end{figure}

We now define the DTI and the TTI when several collision partners are present (see Fig. \ref{Fig_collision_1_3}). We have:

\begin{definition} 
When the subject is interacting with several collision partners at the same time, the subject's global DTI is the minimum of the DTI of all binary encounters. We denote it by $D(w)$ if $w$ is the velocity of the subject. 
\label{def_multiple_interaction}
\end{definition}

\begin{figure}
\begin{center}
\hspace{-2cm}
\input{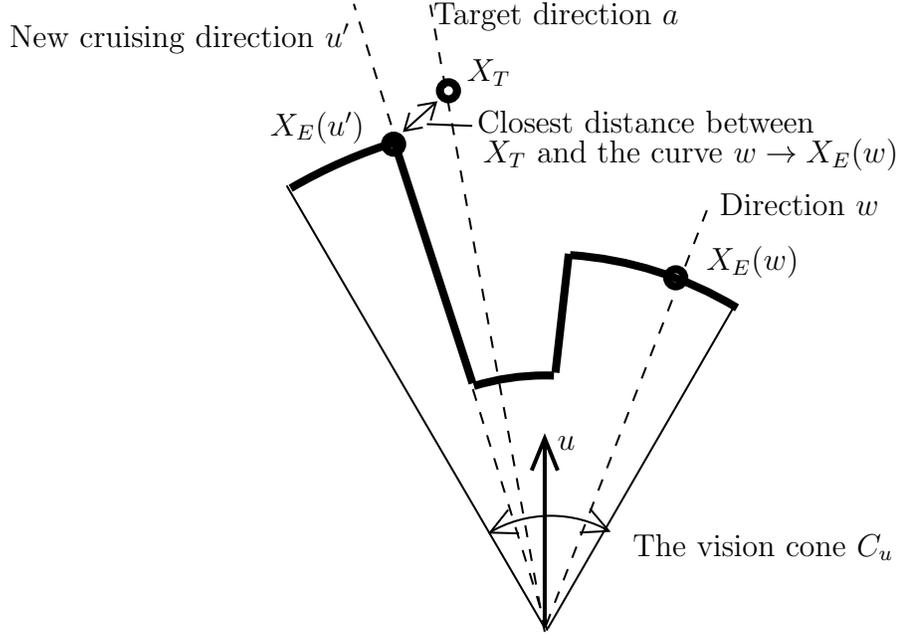}
\caption{Decision-making phase. The new cruising direction $u'$ is chosen such that the estimated point $X_E(u')$ in this direction is the closest to the target point $X_T$. The tested cruising directions $w$ (with associated points $X_E(w)$) are restricted to the vision cone $C_u$ of the pedestrian (where $u$ is the current cruising direction). }
\label{Fig_collision_1_4}
\end{center}
\end{figure}

The decision-making phase consists in changing the current cruising direction $u$ to a new cruising direction $u'$. It is the outcome of an optimization process. From the knowledge of the DTI in each cruising direction, the subject chooses the direction which maximizes the DTI, while keeping his direction of motion close to his target direction. In \cite{Moussaid_etal_PNAS11}, the decision making phase is performed at discrete times separated by equal time intervals $\Delta t$. In this phase, the cruising direction is updated through the following local optimization procedure. Without any obstacle, the subject would choose a target direction $a$ ($a$ is a unit vector of ${\mathbb R}^2$) and cruise with speed $c$. Therefore, after a time interval $\Delta t$, in the absence of obstacles, he would find himself at position $X_T = c \, \Delta t \, a$ (assuming that the origin of the coordinate system is placed at his current position). The point $X_T$ is called the target point. Now, in the presence of obstacles, the subject cruising in the direction $w$ ($w$ being a unit vector in ${\mathbb R}^2$) will estimate impossible to move a distance larger that the DTI. Therefore, choosing the cruising direction $w$ will place the subject at a position $X_E(w) = D(w) w $. The point $X_E(w)$ is the estimated point reached in the direction $w$. The decision making consists in choosing for new cruising direction $u'$ the direction such that the estimated point $X_E(u')$ is the closest to the target point $X_T$, among a set of test directions $w$ belonging to the vision cone $C_u$ about the subject's current direction of motion $u$. Therefore, $u'$ is determined by 
\begin{equation}
u' = \mbox{arg} \min_{w \in C_u} |X_T - X_E(w)|,
\label{eq:min_deci}
\end{equation}
where arg min denotes the point that realizes the minimum. Such a realization of the minimum may not be unique, but we will discard this possibility as non-generic. This decision-making phase is illustrated on Fig. \ref{Fig_collision_1_4}. 

This decision-making rule implicitly states how binary interactions are combined. This combination is not a mere superposition, as in the classical social force model \cite{Helbing_BehavSci91, Helbing_Molnar_PRE95, Helbing_Molnar_SelfOrganization97}, but a highly nonlinear operation involving the solution of an optimization problem. In \cite{Moussaid_etal_PLOSCB12}, it is shown that this model provides a better account of some of the most striking emergence phenomena in crowds, such as spontaneous lane formation in bidirectional motion. 

In the next two sections, we make all these considerations mathematically explicit.

\subsection{Perception phase}
\label{subsec:collision_perception}

\begin{figure}
\begin{center}
\input{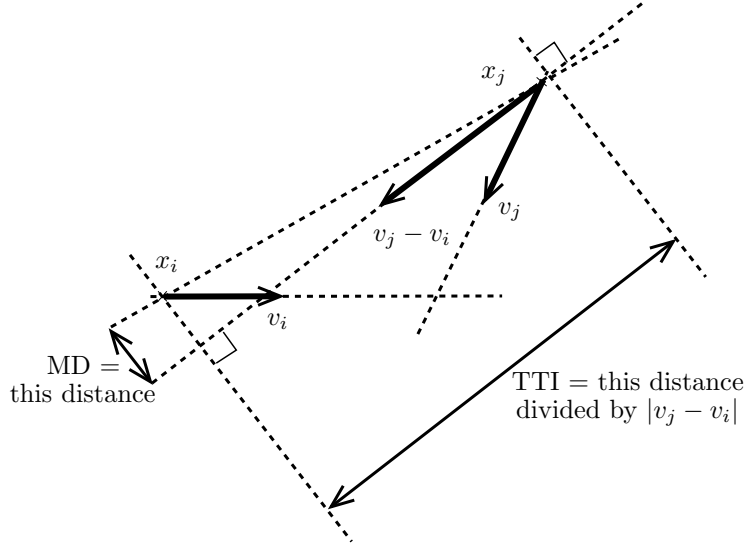}
\caption{Geometry of a collision: The TTI is the projected distance of the two pedestrians on the direction of the relative velocity $v_j-v_i$, divided by the norm of this relative velocity $|v_j-v_i|$. The DTI is the TTI times the velocity of the pedestrian. The MD is the projected distance of the two pedestrians on the normal direction to the relative velocity.  }
\label{Fig_collision_1}
\end{center}
\end{figure}

We consider a pedestrian $i$ located at a position $x_i(t)$, with a velocity $v_i$. He interacts with a collision partner $j$ located at a position $x_j(t)$ which has a velocity $v_j$. Figure \ref{Fig_collision_1} gives a schematic picture of the geometry of the collision. The goal of this section is to compute $\tau_{\mbox{\scriptsize{int}}}$,  $D_{\mbox{\scriptsize{int}}}$, $D_{\mbox{\scriptsize{min}}}$, respectively the TTI, DTI and MD of walker $i$ in his interaction with pedestrian $j$ (see Definition \ref{def_bin_interaction}). 

\begin{lemma}
We have: 
\begin{eqnarray}
\tau_{\mbox{\scriptsize{int}}} &=&  - \frac{ (x_j - x_i) \cdot (v_j - v_i)  }{|v_j - v_i|^2} , \label{eq:tti} \\
D_{\mbox{\scriptsize{int}}} &=& - \frac{ (x_j - x_i) \cdot (v_j - v_i)  }{|v_j - v_i|^2} \, |v_i| ,
\label{eq:dti} \\
D_{\mbox{\scriptsize{min}}}  &=&   \Big( |x_j - x_i|^2 - \big( (x_j - x_i) \cdot \frac{v_j - v_i}{|v_j - v_i|} \big)^2 \Big)^{1/2}.  
\label{eq:min_dist} 
\end{eqnarray}
\label{lem:md_tti}
\end{lemma}

\noindent
{\bf Proof:} The distance $D(t)$ between the two particles at time $t$ is given by 
\begin{eqnarray}
&& \hspace{-1cm}
D^2(t) = |x_j + v_j t - (x_i + v_i t)|^2 \nonumber \\
&& \hspace{-1cm}
= |v_j  - v_i |^2 \Big( t + \frac{(x_j  - x_i) \cdot (v_j  - v_i)}{|v_j  - v_i |^2} \Big)^2 + |x_j  - x_i |^2  - \frac{\big( (x_j  - x_i) \cdot (v_j  - v_i) \big)^2}{|v_j  - v_i |^2}, 
\label{eq:D2}
\end{eqnarray} 
denoting by $x_i$ and $x_j$ the positions of the two particles at time $0$. This quadratic function of time is minimal at the time  
$t=\tau_{\mbox{\scriptsize{int}}}$ given by (\ref{eq:tti}), which gives the value of the TTI. Then, the DTI $D_{\mbox{\scriptsize{int}}}$ of particle $i$ is obviously given by the distance traveled by this particle during the TTI, i.e. $D_{\mbox{\scriptsize{int}}} = \tau _{\mbox{\scriptsize{int}}} \, |v_i|$. This leads to (\ref{eq:dti}). Finally, the MD $D_{\mbox{\scriptsize{min}}}$ is given by the minimal value of (\ref{eq:D2}), i.e. $D_{\mbox{\scriptsize{min}}} = D(\tau_{\mbox{\scriptsize{int}}})$, which leads to (\ref{eq:min_dist}). \endproof

If (\ref{eq:tti}) and (\ref{eq:dti}) give negative
values for the TTI and DTI, it means that there is
no threat of collision in the future times, as pedestrians
are walking away from each other. Therefore, some interaction occurs in the future if and only if $(x_j - x_i) \cdot (v_j - v_i) <0$. Furthermore, if the MD is larger than a certain threshold $R$ identified as the diameter of the individuals, plus a certain safe-keeping distance, the interaction will no longer be perceived as a collision threat. In both cases, the DTI and TTI are set to infinity. With these additional features, we now define the DTI and TTI as 

\begin{definition}
We define: $D_{\mbox{\scriptsize{int}}} $ and $\tau_{\mbox{\scriptsize{int}}}$ as:
\begin{eqnarray}
& & \hspace{-1.5cm}
\left. \begin{array}{l}
\tau_{\mbox{\scriptsize{int}}} =   \frac{ \big| (x_j - x_i) \cdot (v_j - v_i) \big|  }{|v_j - v_i|^2} , \quad 
D_{\mbox{\scriptsize{int}}} = \frac{ \big| (x_j - x_i) \cdot (v_j - v_i) \big|  }{|v_j - v_i|^2}  \, |v_i| , \\
\hspace{0.3cm} \mbox{ if } \, \Big( |x_j - x_i|^2 - \big( (x_j - x_i) \cdot \frac{v_j - v_i}{|v_j - v_i|} \big)^2 \Big)^{1/2} \leq R, \, \mbox{ and } \, (x_j - x_i) \cdot (v_j - v_i) <0 , 
\end{array} \right\}  \label{eq:tti_+} \\
& & \hspace{-1.2cm}
\tau_{\mbox{\scriptsize{int}}} =   + \infty , \quad 
D_{\mbox{\scriptsize{int}}} = + \infty, \hspace{0.4cm} \mbox{ otherwise } .  \label{eq:tti_-} 
\end{eqnarray}
\label{def_Dint_Tauint}
\end{definition}

\subsection{Decision-making phase}
\label{subsec_decision_making_model}

The collision avoidance model of \cite{Moussaid_etal_PNAS11} uses the elements of collision perception reviewed in section \ref{subsec:collision_perception}. In this model, the decision of a new cruising direction taken by the pedestrian reflects the balance between two antagonist goals: collision avoidance on the one hand and maintenance of the target direction on the other hand. The goal of the present paper is to investigate the role of the cruising direction. Consequently, we discard the variations of the cruising speed. We assume that all pedestrians move with constant speed equal to $c$. Therefore, $|v_i|=|v_i|=c$ and we let
$$ v_i = c u_i, \quad v_j = c u_j, \quad |u_i| = |u_j| = 1. $$
This assumption prevents us to take into account one of the features of the model of \cite{Moussaid_etal_PNAS11}, namely that pedestrians slow down or stop in case of very close encounters. This restricts the validity of the present paper to low densities. In the present paper, we will also discard fixed obstacles.

The model follows the lines of \cite{Moussaid_etal_PNAS11}, with some simplifications of the expressions of the collision avoidance response. We assume a time discrete model with time steps $\Delta t$. During this time step, the pedestrian moves according to the vector $v_i \Delta t$. Then, he updates his velocity and adopts a new velocity. For this purpose, he explores all possible velocity directions $w$ and computes the minimum of the DTI with the other pedestrians in the direction $w$. Let us denote by $D_{ij}(w)$ the DTI with pedestrian $j$ assuming that $i$ moves in the direction $w$. If there are no close encounters, we let this quantity be equal to the distance traveled by the pedestrian during $\Delta t $, i.e. $c \Delta t$. In any case, we limit $D_{ij}(w)$ by this quantity. Then, according to (\ref{eq:tti_+}), (\ref{eq:tti_-}), we have: 
\begin{eqnarray}
& & \hspace{-1cm} D_{ij}(w)  =  \left\{ \begin{array}{l} \displaystyle
\min \Big( \, \frac{\big| (x_j - x_i) \cdot (u_j - w) \big|}{|u_j - w|^2}, \, c \Delta t \, \Big) 
\quad \mbox{ if } \quad (x_j - x_i) \cdot (u_j - w) < 0 , \\
\hspace{3.8cm} \mbox{ and } \quad |x_j - x_i|^2 - \big( (x_j - x_i) \cdot \frac{u_j - w}{|u_j - w|} \big)^2 \leq R^2, \\
\vspace{-0.3cm} \mbox{} \\
c \, \Delta t \quad \mbox{ otherwise. }  \end{array} \right. \label{Dij(u)}
\end{eqnarray}
For physical consistency, we should have $R < c \Delta t$, as the typical diameter of a pedestrian should be much less than the distance traveled by a subject between two velocity updates.

Then, we define the minimum $D_i(w)$ of all the DTI by taking the minimum of $D_{ij}(w)$ over all partner pedestrians. The anisotropy of human vision is taken into account by restricting the set of partner pedestrians to those belonging to the vision cone of pedestrian $i$. Introducing a threshold number $\kappa \in [0,1]$, this vision cone is centered at $x_i$ and has half angle $\cos^{-1} \kappa$ about the direction $u_i$. The minimal DTI of the $i$th-pedestrian is therefore defined by :  
\begin{eqnarray}
& & \hspace{-1cm} D_i(w) = \min \, \{ D_{ij}(w), \quad j=1, \ldots, N, \quad j \not = i, \quad \, \frac{x_j - x_i}{|x_j - x_i|} \cdot u_i \geq \kappa \},   
\label{Di(u)}
\end{eqnarray}
where $N$ is the total number of pedestrians. Finally, the new direction of motion $u'_i$ of the $i$-th pedestrian is the direction $w$ that minimizes the distance between the point reached after traveling a distance $D_i(w)$ in this direction and the point reached after traveling a distance $c \Delta t$ in the target direction $a_i$. Therefore, the new direction of motion $u'_i$ is found by solving the following minimization problem: 
\begin{eqnarray}
& & \hspace{-1cm} u_i' = \mbox{arg} \min_{w \in {\mathbb S}^1\, | \, w \cdot u_i \geq \kappa} |D_i(w) w - c \, \Delta t \, a_i|^2,  \label{ui'}
\end{eqnarray}
where again, the test directions $w$ are restricted to the vision cone of pedestrian $i$. We denote by ${\mathbb S}^1$ the set of vectors of ${\mathbb R}^2$ of unit norm. 

\begin{remark}
We note that the minimization problem (\ref{ui'}) is not convex and may have multiple solutions. We will discard this occurrence as non-generic. 
\label{rem:unique}
\end{remark}

\subsection{Summary of the Heuristic-Based IBM model}
\label{subsec:IBM_time_discrete}

We now consider a collection of $N$ point particles with positions $x_i^n \in {\mathbb R}^2$, velocity directions $u_i^n \in {\mathbb S}^1$ at time $t^n = n \Delta t$ and target direction $a_i \in {\mathbb S}^1$, and $i \in \{ 1, \ldots, N \}$. The dynamics is as follows: 
\begin{eqnarray}
& & \hspace{-1cm} x_i^{n+1} = x_i^n + c \, \Delta t \, u_i^n , \label{TDIBM_x} \\
& & \hspace{-1cm} u_i^{n+1} = \mbox{arg} \min_{w \in {\mathbb S}^1\, | \, w \cdot u_i^n \geq \kappa} |D_i^{n+1} (w) w - c \, \Delta t \, a_i|^2 , \label{TDIBM_u} 
\end{eqnarray}
with 
\begin{eqnarray}
& & \hspace{-1.4cm}  
D^{n+1}_i(w) = \min \, \{ D_{ij}^{n+1}(w), \quad j=1, \ldots, N, \quad j \not = i, \quad \, \frac{x_j^{n+1} - x_i^{n+1}}{|x_j^{n+1} - x_i^{n+1}|} \cdot u_i^n \geq \kappa \}, 
\label{Di(n+1)(u)}
\end{eqnarray}
and
\begin{eqnarray}
& & \hspace{-1cm} D_{ij}^{n+1}(w)  =  \left\{ \begin{array}{l} \displaystyle
\min \Big( \, \frac{\big| (x_j^{n+1} - x_i^{n+1}) \cdot (u_j^n - w) \big|}{|u_j^n - w|^2}, \, c \Delta t \, \Big) , \\
\hspace{1.15cm} \mbox{ if } \quad \quad (x_j^{n+1} - x_i^{n+1}) \cdot (u_j^n - w) < 0 , \\
\hspace{1.15cm} \mbox{ and } \quad |x_j^{n+1} - x_i^{n+1}|^2 - \big( (x_j^{n+1} - x_i^{n+1}) \cdot \frac{u_j^n - w}{|u_j^n - w|} \big)^2 \leq R^2, \\
\vspace{-0.3cm} $\mbox{}$ \\
c \, \Delta t \quad \mbox{ otherwise. }  \end{array} \right. \label{Dij(n+1)(u)}
\end{eqnarray}

We now make some comments on the position update rule (\ref{TDIBM_x}). Since the pedestrian can only walk a distance $D(u_i^n)$ in the direction $u_i^n$ before colliding with another pedestrian, it would appear more sensible to use the formula $x_i^n + D(u_i^n) \, u_i^n$. However, in the present model, the pedestrian speed is supposed equal to one. Therefore, this update can only provide the position at an intermediate time $t^n + D(u_i^n)/c < t^{n+1}$. This leads to position updates at different times for the different pedestrians. In order to derive a time continuous model, it is more convenient to keep position updates at constant time-steps, which justifies the choice made in (\ref{TDIBM_x}). In the limit $\Delta t \to 0$ in (\ref{TDIBM_x}) (but keeping $\Delta t$ finite in (\ref{TDIBM_u}), (\ref{Dij(n+1)(u)})), the quantity $c \Delta t$ tends to zero and eventually becomes smaller than $D(u_i^n)$. Then, the objection formulated at the beginning of this paragraph disappears. In the next section, we first propose a time-continuous model based on this limit, and then deduce a mean-field kinetic model from it.

\setcounter{equation}{0}
\section{Mean-field kinetic model}
\label{sec:kinetic}

\subsection{Methodology}
\label{subsec:kinetic_method}

The goal of this section is to propose a time and space continuous kinetic model (KM). With this aim, we first convert the previous time-discrete IBM into a time-continuous one. This conversion consists in replacing the sudden change of the velocity every $\Delta t$ time intervals, by a continuous one. 

The difficulty with writing such a time-continuum model comes from the 'roughness' of the rules of the time-discrete IBM. For this reason, we regularize the time-discrete dynamics in two ways. First, in the perception phase, we replace the distance to the closest encounter by an average distance to the possible encounters in some interaction region. We propose the use of an harmonic average which closely approximates the minimum used in the original model. The use of averages over certain interaction regions is found in many classical swarm models, such as \cite{Aoki_BullJapSocScientFish82, Couzin_etal_JTB02, Degond_Motsch_M3AS08, Gautrais_etal_PlosCB12, Gregoire_Chate_PRL04, Vicsek_etal_PRL95}, but the introduction of harmonic averages is new. Second, in the decision-making phase, the jump to the  direction of motion which maximizes the distance walked towards the target is replaced by a continuous directional change determined by a velocity potential. This supposes that the subjects choose their new cruising direction close to the previous one, which looks realistic.

A final modification of the IBM is to add some noise in the pedestrian  velocity updates. This noise accounts for various effects such as the uncertainties in the estimations of the interaction partner velocities, the variability of the subjects' responses to interactions, the possibility that the subjects react to some unpredicted stimuli, etc. For KM, the introduction of noise in the particle velocity update results in diffusion in velocity space which produces solutions with smooth velocity profiles. This has important consequences for the derivation of CM from KM, as it supports the use of smooth macroscopic closures. Such smooth closures will be at the heart of the VMF closure and of the hydrodynamic limit methodologies which will be described in Sections \ref{subsec:gaussian}  and \ref{subsec:hydro} respectively.

\subsection{Modified time-continuous IBM}
\label{subsec:IBM_time_continuous}

We consider the following time-continuous stochastic model for the pedestrian positions $x_i(t)$ and velocity directions $u_i(t)$: 
\begin{eqnarray}
& & \hspace{-1cm} 
\frac{d x_i}{dt} = c \, u_i(t) , 
\label{IBM_x} \\
& & \hspace{-1cm} 
d u_i = F_i(t) dt + \big( ( \sqrt{2d} \circ dB_i(t) ) \cdot u_i^\bot \big) u_i^\bot, 
\label{IBM_u} 
\end{eqnarray}
where $F_i(t)$ is a force term and $d$ is the noise intensity (supposed uniform among pedestrians). The term $dB_i(t)$ stands for the standard white noise and the symbol '$\circ$' means that the stochastic differential equation must be understood in the Stratonovich sense. The force term $F_i(t)$ is constructed below in such a way that it remains orthogonal to $u_i(t)$, i.e. $F_i(t) \cdot u_i(t) = 0$ and the noise term $\sqrt{2d} \,  dB_i(t)$ is projected onto the orthogonal vector $u_i^\bot$ to $u_i$. These facts, together with the use of the Stratonovich definition of the Stochastic Differential Equation, maintain $u_i$ on the one-dimensional unit sphere i.e. $|u_i(t)| = 1$, provided that $|u_i(0)| = 1$ initially \cite{Hsu_AMS02}. 

The force term is defined as follows. First, we replace the 'min' in (\ref{Di(n+1)(u)}) by an average over neighboring particles located in the $i$-th pedestrian interaction region. We choose an harmonic average, which has the property to give large weights to the small values of the quantity to be averaged. In this way, the harmonic average mimics closely the outcome of the 'min' operation in (\ref{Di(n+1)(u)}). The $i$-th pedestrian interaction region is defined as the angular sector centered at $x_i(t)$, with axis $u_i(t)$, semi-angle $\cos^{-1} \kappa$ and radius $\delta_i(t)$. The set $S_i(t)$ of subjects belonging to the $i$-th pedestrian interaction region is: 
\begin{eqnarray}
& & \hspace{-1cm} S_i(t) = \{ j \in \{1, \ldots , N \} \, \,  | \, \, |x_j(t) - x_i(t)| \leq \delta_i(t) \, \mbox{ and } \, \frac{x_j(t) - x_i(t)}{| x_j(t) - x_i(t) | } \cdot u_i(t) \geq \kappa \} . \label{Si(t)} 
\end{eqnarray}
The angle $\cos^{-1} \kappa$ is the semi-angle of the human vision cone (say typically $\pi/2$, i.e. $\kappa = 0$). The value of $\delta_i(t)$ is linked to the local inter-particle distance and will be estimated later on. The number of elements of $S_i(t)$ is denoted by $\# S_i(t)$. 

We then consider the harmonic average of the elementary DTI with all collision partners in the interaction region: 
\begin{eqnarray}
& & \hspace{-1cm} D_i^{-1}(w,t) = \max \, \Big\{ \frac{1}{\# S_i(t)} \sum_{j \in S_i(t)} D_{ij}^{-1}(w,t) , \frac{1}{L} \, \Big\}, \label{Di(u,t)} 
\end{eqnarray}
where the 'max' has been introduced to bound the average for reasons that will be explained below and where the DTI is defined like in (\ref{Dij(n+1)(u)}): 
\begin{eqnarray}
& & \hspace{-1cm} D_{ij}^{-1}(w,t)  = \left\{ \begin{array}{l} \displaystyle
\min \Big( \, \frac{|u_j(t) - w|^2}{\big| (x_j(t) - x_i(t)) \cdot (u_j(t) - w) \big|}, \, \frac{1}{\ell} \, \Big) , \\
\hspace{1.15cm} \mbox{ if } \quad \quad (x_j(t) - x_i(t)) \cdot (u_j(t) - w) < 0 , \\
\hspace{1.15cm} \mbox{ and } \quad |x_j(t) - x_i(t)|^2 - \big( (x_j(t) - x_i(t)) \cdot \frac{u_j(t) - w}{|u_j(t) - w|} \big)^2 \leq R^2, \\
\vspace{-0.3cm} $\mbox{}$ \\
0 \quad \mbox{ otherwise. }  \end{array} \right. \label{Dij(u,t)}
\end{eqnarray}
The quantity $\ell$ is a lower cut-off for $D_{ij}$ because the elementary DTI can be arbitrarily small. In reality, if the DTI is too small, the pedestrian lowers his velocity or even stops. This feature is not taken into account in the present model, where we only allow directional changes. Therefore, to cope with this situation, our model pedestrian would have to develop very large angular accelerations, which is unrealistic. The parameter $\ell$ is introduced to bound the forces and thus prevent the dynamics to become too singular in this situation. In the situations where the elementary DTI are large (which corresponds to the second alternative of (\ref{Dij(u,t)})), we just set them equal to $\infty$, so that they are not taken into account in the average (\ref{Di(u,t)}), which computes the global DTI. The bound of the global DTI by $c \Delta t$ is realized by the parameter $L$ as described now. 

Indeed, the quantity $L$ stands for the distance walked by the pedestrian between two velocity updates (i.e. $L = c \Delta t$ in the discrete model). Of course, if there are no collision partners (i.e. $S_i(t) = \emptyset$ is the vacuum set), or if the elementary DTI with the available collision partners are large, the pedestrian will be able to walk this distance $L$ without performing a velocity update. Therefore, we bound $D_i$ by $L$ thanks to the 'max' in (\ref{Di(u,t)}). In practice, we have $\ell \sim R \ll L$. Indeed, the lower cut-off for the elementary DTI is of the order of the size of a subject, while the free walking distance between two velocity updates is much larger. 

We now define the $i$-th pedestrian potential function $\Phi_i^t(w)$ for unit vectors $w \in {\mathbb S}^1$ by:
\begin{eqnarray}
& & \hspace{-1cm} \Phi_i^t(w) = \frac{k}{2} |D_i(w,t) w - L \, a_i|^2 . \label{Phi_i} 
\end{eqnarray}
The coefficient $(k L^2)^{-1}$ gives the order of magnitude of the potential and of the force. By (\ref{IBM_u}) and the fact that the velocity $u$ is dimensionless, the force and consequently $(k L^2)^{-1}$ have the physical dimension of a reaction rate. Therefore, we can view the quantity $(k L^2)^{-1}$ as providing the typical magnitude of the pedestrian reaction.  The force $F_i(t)$ is defined by
\begin{eqnarray}
F_i(t) &=& - \nabla_w \Phi_i^t (u_i(t)) \label{eq:F_i_1} \\
&=& - k \,  \Big( \, \big( D_i(u_i(t),t) - L \,  a_i(t) \cdot u_i(t) \big) \,  \nabla_w D_i(u_i(t),t) \nonumber \\
& & \hspace{5cm}
 - D_i(u_i(t),t) \, L \, (a_i(t) \cdot u_i(t)^\bot) u_i(t)^\bot \, \Big). 
\label{eq:F_i_2} 
\end{eqnarray}

We note that gradients of functions defined on ${\mathbb S}^1$ are tangent fields to ${\mathbb S}^1$. Therefore, by formula (\ref{eq:F_i_1}), $F_i(t)$ is orthogonal to $u_i(t)$ as it should. This is reflected in (\ref{eq:F_i_2}). The first term is proportional to $\nabla_w D_i(u_i(t),t)$, while the second one is proportional to $u_i^\bot$, and both are orthogonal to $u_i(t)$. 

Definition (\ref{eq:F_i_1}) reflects the fact that, under the force $F_i$, the pedestrian decreases his potential $\Phi_i^t$. Therefore, the pedestrian turns towards the direction of the local minimum of the attraction basin of $\Phi_i^t$ to which he belongs at time $t$. This rule can be viewed as a local version of the global minimum rule set up by (\ref{TDIBM_u}). Of course, this local minimum may not be the global one expressed by (\ref{TDIBM_u}). However, whether in actual life, a pedestrian spontaneously chooses the global minimum or a local one close to his current direction of motion is not clear. Therefore, to our opinion, this local rule is as legitimate as the global one, until experiments can clarify this point. Of course, the two rules coincide if the local minimum is equal to the global one. So the discrepancy between them may be quite small in practice.

\subsection{Mean-field kinetic model}
\label{subsec:mean_field}

We now introduce a statistical description of the system. Instead of using the 'exact' positions, velocities and preferred directions of pedestrians, we rather describe the system in terms of the probability distribution $f(x,u,a,t)$. Specifically, $f(x,u,a,t) \, dx \, du \, da$ is the probability of finding pedestrians in a small physical volume $dx$ about point $x$, within an angular neighborhood $du$ of velocity direction $u$, and within an angular neighborhood $da$ of preferred direction $a$ at time $t$. We recall that $x \in {\mathbb R}^2$, $u, \, a  \in {\mathbb S}^1$. If there are no interactions between the pedestrians, i.e. if the acceleration term $F$ is due to purely external causes, $f(x,u,a,t)$ can be rigorously proved to satisfy the following Fokker-Planck equation: 
\begin{eqnarray}
&& \hspace{-1cm} \partial_t f + c u \cdot \nabla_x f + \nabla_u \cdot ( Ff) = d \Delta_u f . \label{mf_f}
\end{eqnarray}
This equation is a consequence of Ito's formula of stochastic calculus. The left hand-side is a transport operator. It expresses the material derivative of $f$ in the phase space spanned by $(x,u)$, due to the motion of the particles with velocity $cu$ and acceleration $F$. The right-hand side is a velocity diffusion term which comes from the velocity noise. Let $\theta$  be the angle between $u$ and the first coordinate direction. Then, $ u = (\cos \theta,\sin \theta)$, $u^\bot = (-\sin \theta,\cos \theta)$, and each term of (\ref{mf_f}) is written as follows: 
$$ u \cdot \nabla_x f = \cos \theta \, \partial_{x_1} f + \sin \theta \, \partial_{x_2} f, \quad \nabla_u \cdot ( Ff) = \partial_{\theta} (F_\theta f), \quad \Delta_u f = \partial^2_\theta f, $$
where the force term $ F = F_\theta u^\bot$ is by definition orthogonal to $u$ because $|u|=1$. 

We note that there is no operator acting on the $a$-dependence of $f$. This is because we assume that the target direction $a$ is a quantity attached to the agents which is not changed with time. This assumption could easily be modified to take into account a possible change of the target direction with the motion of the pedestrians. However, even with this simplifying hypothesis, the statistics of target directions has a definite influence on the dynamics through the interaction force described below.  

Here, the acceleration term $F$ is not due to external forces but to interactions between the particles. So the rigorous derivation of (\ref{mf_f}) is more difficult and is left to future work (see e.g. \cite{Bolley_etal_AML12}). The acceleration $F$ is coupled to $f$ through continuous equivalents of (\ref{eq:F_i_1}), and is written: 
\begin{eqnarray}
&& \hspace{-1cm} F(x,u,a,t) = - \nabla_w \Phi_{(x,u,a,t)}(u)  , \label{mf_F}
\end{eqnarray}
where $\Phi_{(x,u,a,t)}(w)$ is the potential of a pedestrian at time $t$ located at $x$ with velocity $u$ and target velocity $a$. The potential is a function of the test direction $w$. It is given by
\begin{eqnarray}
&& \hspace{-1cm} \Phi_{(x,u,a,t)}(w) = \frac{k}{2} |D_{(x,u,t)}(w) w - L \, a|^2  , \label{mf_Phi}
\end{eqnarray}
in terms of the DTI $D_{(x,u,t)}(w)$ of pedestrians located at position $x$ at time $t$ with velocity $u$ in the test direction $w$. 

To compute the DTI, we first define the interaction region of such a pedestrian by:
\begin{eqnarray}
&& \hspace{-1cm} S(x,u,t) = \{ y \in {\mathbb R}^2 \, \, | \, \, |y - x | \leq \delta(x,t) \, , \, \frac{y-x}{|y-x|} \cdot u \geq \kappa \}  , \label{mf_S}
\end{eqnarray}
where $\delta(x,t)$ will be estimated later on. Then, the continuous equivalent of (\ref{Dij(u,t)}) leads to:
\begin{eqnarray}
&& \hspace{-1.5cm} D^{-1}_{(x,u,t)}(w) = \max \, \left\{ \frac{ \int_{y \in S(x,u,t)} \int_{(v,b) \in {\mathbb T}^2} \tilde D^{-1}(y-x,v-w) \, f(y,v,b,t) \, dy \, dv \, db}
{ \int_{y \in S(x,u,t)} \int_{(v,b) \in {\mathbb T}^2} f(y,v,b,t) \, dy \, dv \, db}, \,  \frac{1}{L} \right\}
  . \label{mf_D(x)}
\end{eqnarray}
We have denoted by ${\mathbb T}^2$ the two-dimensional torus ${\mathbb T}^2 = {\mathbb S}^1 \times {\mathbb S}^1$. In (\ref{mf_D(x)}), the quantity $\tilde D(y-x,v-w)$ is the elementary DTI of a pedestrian located at position $x$ and velocity $w$ in the encounter with a particle located at $y$ and having velocity $v$. It is given by:
\begin{eqnarray}
& & \hspace{-1cm} \tilde D^{-1}(y-x,v-w)  =  \left\{ \begin{array}{l} \displaystyle
\min \Big( \, \frac{|v - w|^2}{\big| (y-x) \cdot (v - w) \big|}, \, \frac{1}{\ell} \, \Big) , \\
\hspace{1.15cm} \mbox{ if } \quad \quad (y-x) \cdot (v - w) < 0 , \\
\hspace{1.15cm} \mbox{ and } \quad |y-x|^2 - \big( (y-x) \cdot \frac{v - w}{|v - w|} \big)^2 \leq R^2, \\
0 \quad \mbox{ otherwise. }  \end{array} \right. 
\label{mf_D(x,y)}
\end{eqnarray}
The significance of formulas (\ref{mf_D(x)}) and (\ref{mf_D(x,y)}) and the roles of the parameters $\ell$ and $L$ are the same as in the time-continuous IBM of section \ref{subsec:IBM_time_continuous}.

Collecting (\ref{mf_D(x)}) and (\ref{mf_D(x,y)}) allows us to compute the potential $\Phi_{(x,u,a,t)}(w)$ given by (\ref{mf_Phi}). After computing the gradient, the force (\ref{mf_F}) has the expression: 
\begin{eqnarray}
F(x,u,a,t) &=& - k \,  \Big( \, \big( D_{(x,u,t)}(u) - L \,  a \cdot u \big) \,  \nabla_w D_{(x,u,t)}(u) \nonumber \\
& & \hspace{5cm}
 - D_{(x,u,t)}(u) \, L \, (a \cdot u^\bot) u^\bot \, \Big). 
\label{eq:F_2} 
\end{eqnarray}

Now, we can provide an estimate of $\delta(x,t)$. As the density increases, the mean inter particle distance decreases like $N^{-1/2}(x,t)$ where $N(x,t)$ is the local density: 
\begin{eqnarray}
& & \hspace{-1cm} 
N(x,t) = \int_{(u,a) \in {\mathbb T}^2} f(x,u,a,t) \, du \, da. 
\label{eq:N}
\end{eqnarray}
Therefore, the DTI should decrease in the same proportion. One way to achieve this scaling is by taking $\delta(x,t) \sim N^{-1/2}(x,t)$. Indeed, since $\tilde D(y-x,v-w)$ is of the order of $|y-x|$ (by (\ref{mf_D(x,y)})), the average DTI $D_{(x,u,t)}(w)$ is of the same order.  And since $|y-x| \leq \delta$, we obtain the expected scaling of $D_{(x,u,t)}(w)$ like $N^{-1/2}(x,t)$. In practice, we need to take
\begin{eqnarray}
&& \hspace{-1cm} 
\delta(x,t) = C \, N^{-1/2}(x,t). \label{mf_delta}
\end{eqnarray}
with $C$ sufficiently larger than $1$ to ensure that the estimate (\ref{mf_D(x)}) will take into account enough pedestrians.

Finally, the KM consists of the kinetic equation (\ref{mf_f}), with the acceleration computed through (\ref{eq:F_2}).

\subsection{Mean-field kinetic model: discussion}
\label{subsec:mean_field_discuss}

The mean-field model expresses how the statistical distribution of the pedestrians in position, velocity and target direction evolves with time. This evolution combines a transport operator (left-hand side of (\ref{mf_f})) which describes pedestrian motion towards their target direction and collision avoidance, and a velocity diffusion operator (right-hand side of (\ref{mf_f})), which models velocity uncertainty. The pedestrian speed $c$ is supposed constant because the model focuses on directional changes only. Directional changes are modeled through a force term $F$ (\ref{eq:F_2}), which describes how pedestrians find the best compromise between their target and the necessity of avoiding pedestrians passing by.

The force $F$ is tailored to decrease the potential function $\Phi_{(x,u,t)}(w)$. This potential describes how well the target point is approached when the pedestrian (initially located at position $x$, velocity $u$ and target velocity $a$ at time $t$) moves in direction $w$  (formula (\ref{mf_Phi})). For a set of test velocities $w$, the pedestrian computes his DTI $\tilde D(y-x, v-w)$ with a pedestrian located at $y$ with velocity $v$ (formula (\ref{mf_D(x,y)})) and averages it over all pedestrians located in his vision cone $S(x,u,t)$ (formula (\ref{mf_S})), giving rise to  $D_{(x,u,t)}(w)$ (formula (\ref{mf_D(x)})). This averaged DTI provides him with an estimate of the distance he can move in the direction $w$ and allows him to compute his potential $\Phi_{(x,u,t)}(w)$. Finally, the pedestrian turns to ensure the decay of the potential and to get closer to his goal (formula (\ref{mf_F})). The interaction term is spatially non-local, through (\ref{mf_D(x)}). In the next section, we derive a spatially local approximation of this non-local term.

\subsection{Mean-field kinetic model with local interaction}
\label{subsec:mean_field_local}

If we observe the system at a large distance, the various length scales involved in the interaction terms appear to be small. Therefore, under this assumption, it is legitimate to assume that there exists a small dimensionless quantity $\eta \ll 1$ such that 
\begin{equation} 
\delta = \eta \hat \delta, \quad R = \eta \hat R, \quad L = \eta \hat L, \quad \ell = \eta \hat \ell, 
\label{eq:scaling_local_force_1}
\end{equation}
where all 'hat' quantities are assumed to be ${\mathcal O}(1)$. Simultaneously, we assume that the pedestrian reaction rate remains ${\mathcal O}(1)$. We recall that the pedestrian reaction rate is measured by the coefficient $k L^2$ (see discussion after Eq. (\ref{Phi_i})). This assumption implies that
\begin{equation} 
\eta^2 k =  \hat k={\mathcal O}(1). 
\label{eq:scaling_local_force_2}
\end{equation}

We introduce the change of variables $y = x + \eta \xi$, with $\xi \in {\mathbb R}^2$, in (\ref{mf_D(x)}) and keep only the leading order terms in the expansion in powers of $\eta$. In this scaling Eqs. (\ref{mf_f}) and (\ref{mf_F}) are unchanged, except that all unknowns $f^\eta$, $F^\eta$, $\Phi^\eta$ now depend on $\eta$. Then, the condition $y \in S(x,u,t)$ is equivalent to the condition $\xi \in {\mathcal C}_{u,\kappa,\hat \delta}$, where 
\begin{eqnarray}
&& \hspace{-1cm} 
{\mathcal C}_{u,\kappa,\hat \delta} = \{ \xi \in {\mathbb R}^2 \, \, \,  \big| \, \, \,   |\xi|\leq \hat \delta \quad \mbox{ and } \quad \frac{\xi}{|\xi|} \cdot u \geq \kappa \} . \label{mf_C_local}
\end{eqnarray}
We have 
$$ \tilde D^{-1} (y-x, v-w) = \frac{1}{\eta} \hat D^{-1} (\xi, v-w), $$
with
\begin{eqnarray}
& & \hspace{-1cm} \hat D^{-1}(\xi,v-w)  =  \left\{ \begin{array}{ll} \displaystyle
\min \Big( \, \frac{|v - w|^2}{\big| \xi \cdot (v - w) \big|}, \, \frac{1}{\hat \ell} \, \Big) , & \mbox{ if } \quad \quad \xi \cdot (v - w) < 0 , \\
 & \mbox{ and } \quad |\xi|^2 - \big( \xi \cdot \frac{v - w}{|v - w|} \big)^2 \leq \hat R^2, \\
0 \quad \mbox{ otherwise. }  \end{array} \right. 
\label{mf_D(x,y)_local}
\end{eqnarray}
Consequently, 
$$ D^{-1}_{(x,u,t)}(w) = \frac{1}{\eta} \breve D^{-1}_{(x,u,t)}(w), $$
with 
\begin{eqnarray}
&& \hspace{-1.5cm} \breve D^{-1}_{(x,u,t)}(w) = \max \, \left\{ \frac{ \int_{\xi \in {\mathcal C}_{u,\kappa,\hat \delta}} \int_{(v,b) \in {\mathbb T}^2} \hat D^{-1}(\xi,v-w)  \, f(x+ \eta \xi ,v,b,t) \, d \xi \, dv \, db}
{ \int_{\xi \in {\mathcal C}_{u,\kappa,\hat \delta}} \int_{(v,b) \in {\mathbb T}^2} f(x+ \eta \xi,v,b,t) \, d \xi \, dv \, db} , \,  \frac{1}{\hat L} \right\}.
\label{eq:breveD}
\end{eqnarray}
Finally, we have
$$ \Phi_{(x,u,a,t)}(w) = \frac{\hat k}{2} \big| \breve D_{(x,u,t)}(w) w - \hat L a \big|^2. $$

\begin{figure}[htbp]
\begin{center}
\input{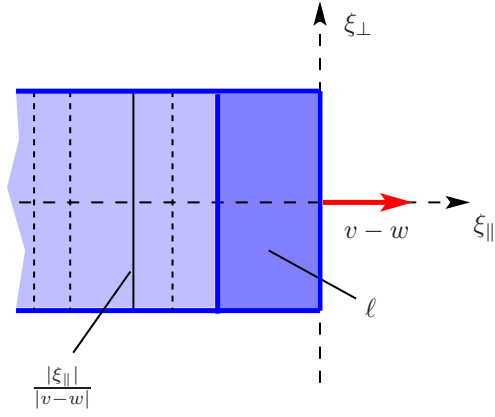}
\caption{The function $\hat D(\xi,v-w)$. For a fixed value of $v-w$ (in red), the parallel and transverse components of $\xi$ are respectively denoted by $\xi_\parallel = \xi \cdot \frac{v-w}{|v-w|}$ and  $\xi_\bot = \xi - \xi_\parallel$. The domain of definition of  $\xi \to \hat D(\xi,v-w)$ (domain where it is finite) is characterized by $\xi_\parallel <0$ and $|\xi_\bot| \leq R$ (see (\ref{mf_D(x,y)_local})) and is the shaded blue area on the figure. The function $\xi \to \hat D(\xi,v-w)$ is constant along all vertical segments and has value $\frac{|\xi_\parallel|}{|v-w|}$ except in the dark blue area where it is constant equal to $\ell$. The transition happens along the line $\xi_\parallel = - \ell |v-w|$ (the vertical blue line on the figure).  }
\label{fig:Dtilde}
\end{center}
\end{figure}

\begin{figure}[htbp]
\begin{center}
\input{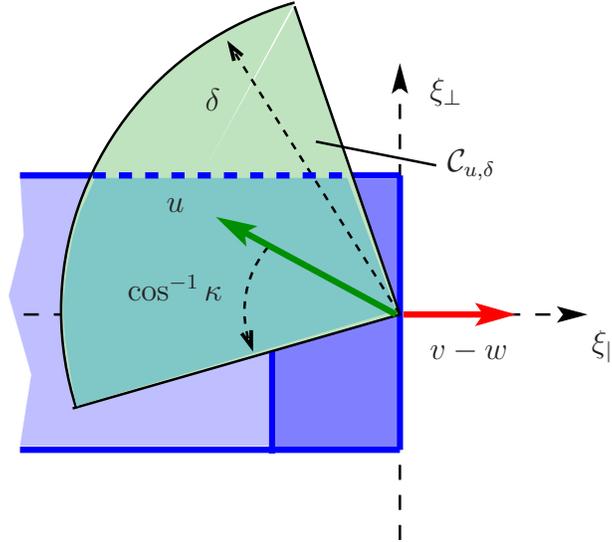}
\caption{The function $\Delta_{\kappa,\delta} (u,v-w)$ is obtained by averaging the function $\xi \to \hat D^{-1}(\xi,v-w)$ (see Fig. \ref{fig:Dtilde}) on the cone ${\mathcal C}_{u,\delta}$ (represented by the green shaded area). The blue-green shaded area is the intersection of the cone ${\mathcal C}_{u,\kappa,\delta}$ and the support of $\xi \to \hat D^{-1}(\xi,v-w)$. It highlights the fact that the function $\Delta_{\kappa,\delta}$ depends on $u \cdot \frac{v-w}{|v-w|}$ and $|v-w|$. }
\label{fig:Delta}
\end{center}
\end{figure}

Since we look for a local approximation, we assume that $\eta \ll 1$. The distribution $f$ is assumed to evolve only on the large scale. Therefore, in the Taylor expansion of (\ref{eq:breveD}) with respect to $\eta$, we may keep only the leading order term and neglect the higher order ones. As a result of this approximation, $\xi$ disappears from the arguments of the function $f$ in both the numerator and denominator. The integration with respect to $\xi$ can thus be performed beforehand, leading to the quantity $\Delta^{-1}_{\kappa, \hat \delta} (u,v-w)$ defined by
\begin{eqnarray}
&& \hspace{-1cm} 
\Delta^{-1}_{\kappa, \hat \delta} (u,v-w) = \frac{1}{\mbox{Area}({\mathcal C}_{u,\kappa,\hat \delta})} \int_{{\mathcal C}_{u,\kappa,\hat \delta}} \hat D^{-1} (\xi, v-w) \, d \xi , 
\label{mf_Delta_local}
\end{eqnarray}
and $\mbox{Area}({\mathcal C}_{u,\kappa,\hat \delta})$ is the two-dimensional area of ${\mathcal C}_{u,\kappa,\hat \delta}$. 
This leads to the following expression of $\breve D$, dropping all the hats for simplicity:
\begin{eqnarray}
&& \hspace{-1cm} \breve D^{-1}_{(x,u,t)}(w) = \max \, \left\{ \frac{ \int_{(v,b) \in {\mathbb T}^2} \Delta^{-1}_{\kappa, \delta(x,t)} (u,v-w) \, f(x,v,b,t) \, dv \, db}
{ \int_{(v,b) \in {\mathbb T}^2} f(x,v,b,t) \, dv \, db} , \,  \frac{1}{L} \right\}
  . \label{mf_D(x)_local}
\end{eqnarray}
Again, $\delta(x,t)$ is linked to the total density through (\ref{mf_delta}). Graphical representations of $\hat D^{-1}$ and $\Delta_{\kappa, \delta}$ can be found in Figs. \ref{fig:Dtilde} and \ref{fig:Delta} respectively. They illustrate that the function $\Delta_{\kappa, \delta}$ only depends on $u \cdot \frac{v-w}{|v-w|}$ and $|v-w|$ (i.e. two real variables) while a general function of $(u,v-w)$ depends on a vector of ${\mathbb S}^1$ and a vector of ${\mathbb R}^2$, i.e. three real variables. This is due to the fact that $\hat D$ itself only depends on $\xi \cdot (v - w)$ and $|v - w|$. The function $\Delta_{\kappa, \delta}$ can be numerically computed a priori. 

\begin{figure}[htbp]
\begin{center}
\input{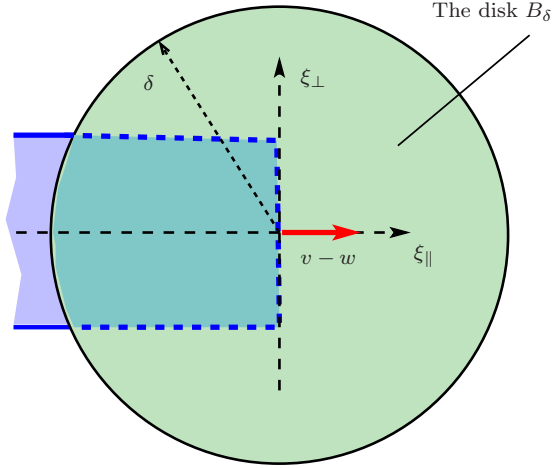}
\caption{Case $\kappa = -1$ (i.e. there is no blind zone: all collision partners in the disk $B_\delta$ of radius $\delta$ are taken into account). The function $\Delta_{\delta} (|v-w|)$ is obtained by averaging the function $\xi \to \hat D^{-1}(\xi,v-w)$ (see Fig. \ref{fig:Dtilde}) on the disk $B_\delta$ (represented by the green shaded area). The blue-green shaded area is the intersection of $B_\delta$ and the support of $\xi \to \hat D^{-1}(\xi,v-w)$. The function $\Delta_{\delta}$ depends only on $|v-w|$.}
\label{fig:Delta_0}
\end{center}
\end{figure}

We note that the expression of $\Delta_{\kappa, \delta}$ simplifies in the special case $\kappa = -1$. In this case, there is no blind zone: all their collision partners in the disk 
\begin{equation}
B_\delta  = \{ \xi \in {\mathbb R}^2  \,  \big| \,   |\xi|\leq \delta \} ,
\label{eq:Bdelta}
\end{equation}
are taken into account by the pedestrians. Consequently, the averaging (\ref{mf_Delta_local}) is over the whole disk $B_\delta$ (see Fig. \ref{fig:Delta_0}). The dependence of $\Delta_{-1, \delta} $ upon $u$ disappears. The resulting function, denoted by $\Delta_{\delta} (|v-w|)$, is given by:
\begin{eqnarray}
&& \hspace{-1cm} 
\Delta^{-1}_{\delta} (|v-w|) = \frac{1}{\mbox{Area}(B_{\delta})} \int_{B_{\delta}} \hat D^{-1} (\xi, v-w) \, d \xi . 
\label{mf_Delta_local_iso}
\end{eqnarray}
Additionally, it is an elementary matter to remark that when $|v-w| \to 0$, we have $\Delta^{-1}_{\delta} (|v-w|) \sim |v-w| \, \ln \frac{|v-w|}{\ell}$. Another consequence of this simplification is that the potential $\Phi$ does not depend on $u$. It can be simply written $\Phi_{x,a,t}(w)$. This simplification will be exploited in the hydrodynamic limit (see section \ref{subsec:hydro}). 

We summarize this section: due to the assumption that $f$ evolves on the large scale only and thus can be taken constant in the interaction region of a given pedestrian, the interaction force only depends on $f$ at that location. This local approximation scaling (\ref{eq:scaling_local_force_1}), (\ref{eq:scaling_local_force_2}) leads to the kinetic model (\ref{mf_f}) with a local evaluation of the force. The force is still computed from the potential (\ref{mf_Phi}) through (\ref{mf_F}). However, the evaluation of the DTI is now given by a local velocity average (\ref{mf_D(x)_local}), where the velocity convolution kernel $\Delta^{-1}_{\kappa, \delta} (u,v-w)$ can be analytically computed. In the simpler case where there is no blind zone, the kernel reduces to a function of $|v-w|$ only, and the potential $\Phi_{x,a,t}(w)$ does not depend on $u$.

\setcounter{equation}{0}
\section{Macroscopic model}
\label{sec:macro}

\subsection{Necessity of a closure Ansatz}
\label{subsec:macro_closure}

Macroscopic models are obtained by taking averages of functions of the particle velocity $u$ over the distribution function $f(x,u,a,t)$. The resulting macroscopic quantities are e.g. the density $\rho(x,a,t)$ or the mean velocity $U(x,a,t)$ of pedestrians at position $x$ with target direction $a$ at time $t$:  
\begin{eqnarray}
& & \hspace{-1cm} 
\rho(x,a,t) = \int_{u \in {\mathbb S}^1} f(x,u,a,t) \, du, \label{eq:moments1} \\
& & \hspace{-1cm} 
U(x,a,t) = \frac{1}{\rho(x,a,t)} \, \int_{u \in {\mathbb S}^1} f(x,u,a,t) \, u \, du.
\label{eq:moments2}
\end{eqnarray}
It is necessary to keep the dependence of the macroscopic quantities over the target direction $a$. Indeed, in general, the statistics of the target directions is not known or may change from situation to situation. In situations where the statistics of target directions is known and does not change with time, it is also possible to introduce the total density $N(x,t)$ (already met at (\ref{eq:N})) and the average velocity $V(x,t)$ of the pedestrians at position $x$ and time $t$, irrespective of their target direction. The latter is defined by: 
\begin{eqnarray*}
\hspace{-1cm} 
V(x,t) = \frac{1}{N(x,t)} \, \int_{(u,a) \in {\mathbb T}^2} f(x,u,a,t) \, u \, du \, da.
\end{eqnarray*}
In the present work, we will only consider CM which retain the statistics of target directions. 

To pass from the KM (\ref{mf_f}) to a CM for the quantities $\rho$ and $U$, the most direct method is the moment method. It consists in integrating the kinetic equation (\ref{mf_f}) with respect to the particle velocity $u$, after pre-multiplication by polynomial functions of $u$. Unfortunately, in general, this method does not yield a closed model for $\rho$ and $U$ because higher order moments (i.e. integrals of higher order polynomials of $u$) may be involved in the resulting system. These higher moments need to be expressed in terms of $\rho$ and $U$ through a suitable closure relation. Closure relations are usually provided through an Ansatz which expresses $f$ itself as a function of $\rho$ and $U$. The validity of this Ansatz is subject to caution. When dissipative phenomena are present, such as in gas dynamics, it is possible to justify it through the hydrodynamic limit (see a review on these questions e.g. in \cite{Degond_review_Birkhauser_03}). Here, the hydrodynamic limit can be developed solely in the special case where the interaction is local (as in section \ref{subsec:mean_field_local}) and in the absence of any blind zone behind the subject. We will first propose two other closure methodologies which apply to general cases, but which cannot be justified by a hydrodynamic procedure.

The first closure scheme, referred to as the 'monokinetic closure', is developed in section \ref{subsec_monokinetic}. It is valid when there is rigorously no noise (i.e. no uncertainty in the pedestrian velocities). It postulates a monokinetic distribution function: in the neighborhood of a given location $x$ at time $t$, all pedestrians having the same target direction $a$ have the same velocity $U(x,a,t)$. In other words, in this neighborhood, the statistics of possible velocities is given by a Dirac delta in the velocity variable, located at $U(x,a,t)$. The resulting CM belongs to the class of second-order models of traffic: it involves two balance equations for the mass and momentum densities respectively and bears analogies with pressureless gas dynamics models \cite{Bouchut_AdvKinTheoryComputing94, Bouchut_James_CPDE99}. These models have somehow unpleasant features, such as the possible formation of mass concentrations. 

For this reason, a second closure scheme, referred to as the 'VMF closure', is proposed in section \ref{subsec:gaussian}. The model supposes that some noise is involved in the pedestrian velocities; this is indeed more realistic than the zero-noise assumption of the previous closure. The distribution of velocities is supposed to be a von Mises-Fisher (VMF) distribution. The VMF distribution is a natural extension of the standard Gaussian distribution for random variables defined on the sphere \cite{Watson_JAP82}. Like in the monokinetic closure scheme, the resulting CM belongs to the class of second-order models of traffic but the form of the momentum equation has not been previously found anywhere else. 

Finally, in section \ref{subsec:hydro}, we develop the hydrodynamic limit in the special case of a local interaction with no blind zone. The hydrodynamic limit supposes that, for a pedestrian, the process of turning towards the velocity which minimizes the potential $\Phi$ is very short. Therefore, the velocity distribution can be approximated by an equilibrium which reflects the instantaneous equilibrium between the turning process and the noise. Such a  distribution, which will be our closure Ansatz, is called a 'Local Thermodynamical Equilibrium' (LTE), by analogy to the standard terminology of statistical mechanics. The LTE is very peaked around the velocity which minimizes the potential, with some spread due to the noise. An important point is that, while the LTE depends on the potential, the potential also depends on the LTE through the definition of the DTI. Therefore, the allowed DTI are determined by a fixed point equation. The resulting LTE can be interpreted as a Nash equilibrium of a game consisting for the pedestrians in finding the best compromise between reaching their target and avoiding collisions with other pedestrians. The framework for a game-theoretic interpretation of LTE can be found in \cite{Degond_etal_arXiv:1212.6130}. The resulting model is a first-order model, in the traffic terminology sense. It consists of a conservation equation for the mass density, while the mass flux is determined functionally from the LTE, i.e. from the DTI that have been found by solving the consistency equation.

\subsection{Monokinetic closure}
\label{subsec_monokinetic}

\subsubsection{Monokinetic closure: derivation}
\label{subsubsec_monokinetic_derivation}

\begin{figure}[htbp]
\begin{center}
\input{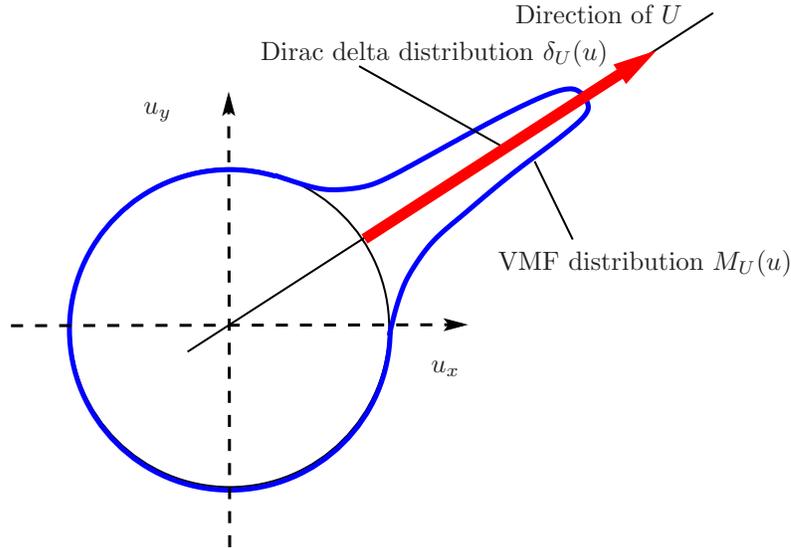}
\caption{The VMF distribution (in blue) and the Dirac delta distribution (in red) as functions of $u$ in polar coordinates. The direction of the mean velocity $U$ is given by the black semi-line. The width of the VMF distribution about its maximum $u=U$ is a function of $\beta(|U|)$. In both cases, the mean velocity $U$ is a function of $(x,a,t)$ determined by the fluid model. We have $|U|=1$ in the Dirac delta distribution case and $|U|<1$ in the VMF case. }
\label{fig:VMF}
\end{center}
\end{figure}

In this section, in order to derive a macroscopic model, we assume a monokinetic distribution function. For the monokinetic assumption to be valid, we need to  remove the noise term, and consider the following equation: 
\begin{eqnarray}
& & \hspace{-1cm} \partial_t f + c u \cdot \nabla_x f + \nabla_u \cdot ( Ff) = 0 , \label{mf_f_nn}
\end{eqnarray}
coupled to (\ref{mf_F}). The monokinetic closure consists of the Ansatz:
\begin{eqnarray}
& & \hspace{-1cm} f(x,u,a,t) = \rho(x,a,t) \delta_{U(x,a,t)}(u), \label{mk_closure}
\end{eqnarray}
where $\delta_{U}(u)$ is the Dirac delta located at $U$ (see a graphical representation at Fig. \ref{fig:VMF} (red arrow)). Note that, by definition, $U(x,a,t) \in {\mathbb S}^1$ i.e. is a vector of norm $1$. 
This Ansatz means that there is only one definite velocity $U(x,a,t)$ at any given point $x$, time $t$, for any preferred direction $a$. It is easily shown \cite{Degond_review_Birkhauser_03} that the distribution (\ref{mk_closure}) is an {\em exact} solution of (\ref{mf_f_nn}) provided that $\rho$ and $U$ satisfy the following set of macroscopic equations: 
\begin{eqnarray}
& & \hspace{-1cm} \partial_t \rho + \nabla_x \cdot (c \rho U) = 0 , \label{mk_rho} \\
& & \hspace{-1cm} \partial_t U + c U \cdot \nabla_x  U = \bar F(x,a,t) , \label{mk_U} 
\end{eqnarray}
with $\bar F(x,a,t) = F(x,U(x,a,t),a,t)$ and $F$ given by (\ref{mf_F}). In other words, 
\begin{eqnarray}
& & \hspace{-1cm}  \bar F(x,a,t) = - \nabla_w \bar \Phi_{(x,a,t)}(U(x,a,t)),\quad \mbox{ with } \quad \bar \Phi_{(x,a,t)}(w) = \Phi_{(x,U(x,a,t),a,t)}(w), 
\label{mk_F}
\end{eqnarray}
and $\Phi$ given by (\ref{mf_Phi}). The potential $\bar \Phi_{(x,a,t)}(w)$ can be written:
\begin{eqnarray}
&& \hspace{-1cm} \bar \Phi_{(x,a,t)}(w) = \frac{k}{2} |\bar D_{(x,a,t)}(w) w - L \, a|^2  , \label{mk_Phi}
\end{eqnarray}
where $\bar D_{(x,a,t)}(w) = D (x,U(x,a,t),w,t)$ is given by:
\begin{eqnarray}
&& \hspace{-1cm} \bar D^{-1}_{(x,a,t)}(w) = \max \left\{ \frac{ \int_{y \in \bar S(x,a,t)} \, \int_{b \in {\mathbb S}^1} \tilde D^{-1}(y-x,U(y,b,t)-w) \, \rho(y,b,t) \, dy \, db}{\int_{y \in \bar S(x,a,t)} \, \int_{b \in {\mathbb S}^1} \rho(y,b,t) \, dy \, db}, \, \frac{1}{L} \right\}  \label{mk_D(x)}
\end{eqnarray}
with 
\begin{eqnarray}
&& \hspace{-1cm} 
\bar S(x,a,t) = \{ y \in {\mathbb R}^2 \, \, | \, \, |y-x|\leq \delta(x,t), \, \frac{y-x}{|y-x|} \cdot U(x,a,t) \geq \kappa \} , 
\label{mk_S}
\end{eqnarray}
and the functions $\tilde D(y-x, v-w)$ and $\delta(x,t)$ still given by (\ref{mf_D(x,y)}) and (\ref{mf_delta}). 

We note that, by definition (\ref{mk_F}), $\bar F$ is orthogonal to $U$. Then, multiplying (\ref{mk_U}) scalarly by $U$ and using that the operator $\partial_t + c U \cdot \nabla_x$ is a derivative, we get: 
$$ (\partial_t + c U \cdot \nabla_x) (|U|^2) = 0. $$
Therefore, the constraint $|U|=1$ is satisfied at any time provided it is satisfied at time $t=0$.

\subsubsection{Monokinetic closure: discussion}
\label{subsubsec_monokinetic_discussion}

The model expresses the conservation of mass (\ref{mk_rho}) and evolution of velocity (\ref{mk_U}). The mass conservation equation (\ref{mk_rho}) takes the form of a classical continuity equation and expresses that the rate of change of $\rho$ in any arbitrary small volume is solely due to the mass flow across the boundary of this volume. The velocity equation (\ref{mk_U}) expresses that the rate of change of $U$ along the flow lines (i.e. the left-hand side of (\ref{mk_U}) which takes the form of a material derivative of $U$) is proportional to the force $\bar F$ exerted on these particles. The target direction $a$ does not explicitly appear in (\ref{mk_rho}) and (\ref{mk_U}) except through this force term which couples all target directions altogether. 

The force term (\ref{mk_F}) describes how the bulk fluid velocity $U$ changes in time: it tends to decrease the potential (\ref{mk_Phi}), whose minima express the best satisfaction of the target direction while avoiding collisions. The potential is computed as follows. By the monokinetic assumption (\ref{mk_closure}), all pedestrians within a given fluid element which have the same target direction $a$ also have the same velocity $U(x, a, t)$. Then, the elementary DTI of these pedestrians with pedestrians located at $y$ and having target direction $b$ is computed. Again, by the monokinetic assumption (\ref{mk_closure}), these particles have velocities $U(y,b,t)$. Therefore, this elementary DTI computed with the test velocity update $w$ is given by $\bar D(y-x,U(y,b,t)-w)$, where 
$\tilde D(y-x,v-w)$ is the elementary DTI (see \ref{mf_D(x,y)}). Then, these elementary DTI are averaged over particles located in the cone of vision of $x$, defined at (\ref{mk_S}). This provides the averaged DTI $\bar D_{(x,a,t)}(w)$ with test velocity update $w$ (formula \ref{mk_D(x)}). The average DTI is used to construct the potential $\Phi_{(x,a,t)}(w)$ which expresses how far the pedestrian is from his target point when walking in the direction $w$. 

The expression of the force term is non-local in space: it involves the complex average (\ref{mk_D(x)}) over a neighborhood of the point where the force evaluation is made. This non-locality expresses the ability of the pedestrian to anticipate the likelihood of a collision with the neighbors. However, a local version of this model can be designed, based on the local version of the kinetic model of section \ref{subsec:mean_field_local}. This local version is simply obtained by replacing (\ref{mk_D(x)}) by its local version issued from (\ref{mf_D(x)_local}). It leads to 
\begin{eqnarray}
&& \hspace{-1cm} D^{-1}_{(x,a,t)}(w) = \max \, \left\{ \frac{ \int_{b \in {\mathbb S}^1} \Delta^{-1}_{\kappa, \delta(x,t)} (U(x,a,t),U(x,b,t)-w) \, \rho(x,b,t) \, db}
{ \int_{b \in {\mathbb S}^1} \rho(x,b,t) \, db} , \,  \frac{1}{L} \right\}
  , \label{mk_D(x)_local}
\end{eqnarray}
with the function $\Delta_{\kappa, \delta} (u,v-w) $ defined by (\ref{mf_Delta_local}). As seen in section \ref{subsec:mean_field_local}, in the special case $\kappa = -1$ (i.e. there is no blind zone behind the subject), the function $\Delta_{\kappa, \delta}(u,v-w) $ is replaced by $\Delta_{\delta}(|v-w|) $, which does not depend on $u$. In this case, the DTI does not depend on $a$ and is given by the expression: 
\begin{eqnarray}
&& \hspace{-1cm} D^{-1}_{(x,t)}(w) = \max \, \left\{ \frac{ \int_{b \in {\mathbb S}^1} \Delta^{-1}_{\delta(x,t)} (|U(x,b,t)-w|) \, \rho(x,b,t) \, db}
{ \int_{b \in {\mathbb S}^1} \rho(x,b,t) \, db} , \,  \frac{1}{L} \right\}
  , \label{mk_D(x)_local_2}
\end{eqnarray}
In all these cases, the evaluation of the force still requires an integration in the target direction variable.

Apart from the complex expression of the force term, this model belongs to the class of pressureless gas dynamics models \cite{Bouchut_AdvKinTheoryComputing94, Bouchut_James_CPDE99}. Such models have some pathologies: they do not guarantee that the monokinetic closure assumption (see Fig. \ref{fig:VMF} (red arrow)) is preserved in time. Specifically, particle trajectories with same target direction $a$ but initially located at different positions $x_0$ and $x'_0$ can meet at later times. This results in the appearance of a  discontinuity of $U$ (because the two meeting particle trajectories may have different velocities) and the blow-up of $\rho$. This classical phenomenon is similar to the appearance of caustics in geometrical optics. The non-local force term $\bar F(x,a,t)$ at the right-hand side of (\ref{mk_U}) is likely to be too weak to repel the trajectories at close encounters. To  prevent this blow-up, it is necessary to introduce some kind of 'internal energy'. This is the motivation for the VMF closure below.

\subsection{VMF closure}
\label{subsec:gaussian}

\subsubsection{VMF closure: derivation}
\label{subsubsec_gaussian_derivation}

In the previous section, it was not needed to take the moments of the noiseless kinetic equation (\ref{mf_f_nn}), since the monokinetic Ansatz (\ref{mk_closure}) provides an exact solution of it. Here, we will consider an Ansatz which is a priori not a solution of the kinetic equation (\ref{mf_f}) (but which hopefully is close to one). But as a counterpart, we will be able to take into account the noise term. With this aim, we take the first two moments of (\ref{mf_f}). 

First, integrating (\ref{mf_f}) with respect to $u$ leads to the mass conservation equation in the same form as previously: 
\begin{eqnarray}
& & \hspace{-1cm} \partial_t \rho + \nabla_x \cdot (c \rho U) = 0 . \label{gauss_rho}  
\end{eqnarray}
Indeed, the Fokker-Planck equation (\ref{mf_f}) is of the form 
\begin{eqnarray}
& & \hspace{-1cm} \partial_t f + \nabla_x \cdot (c u f ) = \nabla_u \cdot (A_1 + A_2) ,   
\label{eq:mf}
\end{eqnarray}
where
\begin{eqnarray}
& & \hspace{-1cm} A_1(u) = - Ff, \quad A_2(u) =  d \nabla_u f,   
\label{eq:mf_rhs}
\end{eqnarray}
are tangent vector fields to ${\mathbb S}^1$. Therefore, thanks to Stokes's formula, 
$$ \int_{u \in {\mathbb S}^1} \nabla_u \cdot (A_1 + A_2) \, du = 0. $$
Since $u$, $x$ and $t$ are independent variables, integration with respect to $u$ commutes with derivation with respect to $t$ and $x$ and, with the definitions (\ref{eq:moments1}), (\ref{eq:moments2}), the integral of the left hand-side of (\ref{eq:mf}) leads to the left-hand side of (\ref{gauss_rho}). 

We now turn to the equation for $\rho U$. Multiplying (\ref{eq:mf}) by $u$ and integrating with respect to $u$ leads to 
\begin{eqnarray}
& & \hspace{-1cm} 
\partial_t (\rho U) + \nabla_x \cdot ( c \Sigma ) = \int_{u \in {\mathbb S}^1} \nabla_u \cdot (A_1 + A_2) \, u \,  du  , \label{gauss_U}  
\end{eqnarray}
with the $2 \times 2$ tensor $\Sigma$ defined by
$$ \Sigma = \int_{u \in {\mathbb S}^1} f \, u \otimes u \, du, $$
and $u \otimes u$ is a matrix of components $(u \otimes u)_{ij} = u_i \, u_j$. Thanks to Stokes's formula, we have:
\begin{eqnarray} 
& & \hspace{-1cm} 
\int_{u \in {\mathbb S}^1} \nabla_u \cdot (A_1 + A_2) \, u \, du = - \int_{u \in {\mathbb S}^1} (A_1 + A_2) \, du. 
\label{eq:A1A2}
\end{eqnarray}
In particular, for $A_2$, applying Stokes's formula once more, we get 
\begin{eqnarray*}
& & \hspace{-1cm} \int_{u \in {\mathbb S}^1} (\nabla_u \cdot A_2) \, u \, du = - d \int_{u \in {\mathbb S}^1} \nabla_u f \, du = - d \int_{u \in {\mathbb S}^1} f \, u  \, du = - d \rho U. 
\end{eqnarray*}
The other integrals, namely $\Sigma$ and that related to $A_1$ cannot be expressed analytically from $\rho$ and $U$. In particular, $\Sigma$ involves second order moments of $f$ with respect to $u$. To proceed, we need a closure assumption. 

By analogy with gas dynamics, we assume that the velocity distribution is a von Mises-Fisher (VMF) distribution about the mean direction of $U$. The VMF distribution is what generalizes the Gaussian measure to the circle and more generally, to spheres \cite{Watson_JAP82}. In the present context, its expression is given by 
\begin{eqnarray}
& & \hspace{-1cm} 
M_U(u) = \frac{1}{Z} \exp\{ \beta \, (u \cdot \Omega) \}, \quad \Omega = \frac{U}{|U|}, 
\label{eq:VMF_norm}
\end{eqnarray}
The quantity $Z$ is a normalizing constant such that $M_U$ is a probability density on ${\mathbb S}^1$: 
\begin{eqnarray}
& & \hspace{-1cm} 
\int_{u \in {\mathbb S}^1} M_U \, du = 1. 
\label{eq:VMF_norm_cnd}
\end{eqnarray}
Then we have 
\begin{eqnarray*}
Z =\int_{u \in {\mathbb S}^1} \exp\{ \beta \, (u \cdot \Omega) \} \, du .
\end{eqnarray*}
Introducing the angle $\theta = \widehat{(\Omega, u)}$, we can write:
\begin{eqnarray}
Z =\int_0^{2 \pi} \exp\{ \beta \, \cos \theta \} \, d \theta = 2 \pi I_0(\beta) .
\label{eq:Z}
\end{eqnarray}
We recall that  $I_k(x)$ denotes the modified Bessel function of the first kind:
$$ I_k(x) = \frac{1}{\pi} \int_0^{\pi} \exp\{ x \, \cos \theta \} \, \cos (k \, \theta) \, d \theta, \quad \forall x \in {\mathbb R}, \quad  \forall k \in {\mathbb N}.  $$
The constant $\beta$ plays the role of an inverse temperature: If $\beta$ is large, $M_U(u)$ is extremely peaked in the direction $u = \Omega$ while if $\beta$ is small, $M_U(u)$ is almost isotropic. It will be determined later on. A graphical representation of the VMF distribution $M_{U}(u)$ as a function of $u$ in polar coordinates is given at Fig. \ref{fig:VMF} (blue curve). The function $M_{U}(u)$ is maximal at $u=U$. Its graphical representation is bell-shaped in a neighborhood of this maximum with a width roughly proportional to $\sqrt{\beta}$. 

Now, we assume the VMF closure Ansatz, namely that $f$ is proportional to $M_U$ and is written:
\begin{eqnarray}
& & \hspace{-1cm} 
f(x,u,a,t) = \rho(x,a,t) M_{U(x,a,t)} (u), 
\label{eq:VMF}
\end{eqnarray}
where $\rho(x,a,t)$ and $U(x,a,t)$ are the moments (\ref{eq:moments1}) and (\ref{eq:moments2}) of $f$. That (\ref{eq:VMF}) satisfies (\ref{eq:moments1}) is obvious in view of the normalization condition (\ref{eq:VMF_norm_cnd}). However, that it satisfies (\ref{eq:moments2}) requires a relation between $\beta$ and $|U|$ as we see now. Eq. (\ref{eq:moments2}) requires that 
\begin{eqnarray}
\int_{u \in {\mathbb S}^1} \frac{1}{Z} \exp\{ \beta \, (u \cdot \Omega) \} \, u \, du = U.
\label{eq:compa}
\end{eqnarray}
We decompose $u$ onto the direction spanned by $\Omega$ and its orthogonal: 
$$ u = (u \cdot \Omega) \Omega + (u \cdot \Omega^\bot) \Omega^\bot. $$
Inserting this decomposition into (\ref{eq:compa}) and noting that the first term is an even function of $\theta$ and the second one, an odd function of $\theta$, we find that (\ref{eq:compa}) is equivalent to:
\begin{eqnarray*}
\int_{u \in {\mathbb S}^1} \frac{1}{Z} \exp\{ \beta \, (u \cdot \Omega) \} \, (u \cdot \Omega) \, du = |U|.
\end{eqnarray*}
or equivalently, to:
\begin{eqnarray*}
\frac{1}{Z} \int_0^{2 \pi} \exp\{ \beta \, \cos \theta \} \, \cos \theta \, d \theta = |U| .
\end{eqnarray*}
This equation can be put in the form
\begin{eqnarray}
\frac{I_1(\beta)}{I_0(\beta)} =  |U|.
\label{eq:compa2}
\end{eqnarray}
The left-hand side is a monotonically increasing function of $\beta \in [0,\infty)$ onto $[0,1)$ \cite{Degond_etal_arXiv:1109.2404}. Therefore, as long as $|U|<1$, there exists a unique $\beta$ such that (\ref{eq:compa2}) holds. We will denote this root by $\beta = \beta(|U|)$. 

That $|U|$ should be less than $1$ is a consequence of the kinetic model. Since the microscopic velocities $u$ are normalized, the norm of their average over the probability $M_U$ should be less than $1$. The parameter $|U|$ is the classical order parameter of nematic materials (see applications to swarm dynamics in \cite{Gregoire_Chate_PRL04, Degond_etal_arXiv:1109.2404, Vicsek_etal_PRL95}). If  $|U|$ is close to $0$, then, $\beta$ is close to $0$ and the distribution is almost isotropic. This indicates a disordered state, with microscopic velocities $u$ pointing in all possible directions, so that their average almost cancels out. On the other hand, if $|U|$ is close to $1$, then $\beta$ is very large and the distribution is strongly peaked about the mean velocity $\Omega$. This is the case where almost all microscopic velocities are pointing in the direction of $\Omega$. 

Since $Z$ depends on $\beta$ through (\ref{eq:Z}) and $\beta$ depends on $|U|$ through (\ref{eq:compa2}), we will now write $\beta = \beta (|U|)$, $Z = Z(|U|)$. The VMF distribution (\ref{eq:VMF}) is now written (omitting the dependences of $\rho$ and $U$ upon $(x,a,t)$  for clarity):
\begin{eqnarray}
& & \hspace{-1cm} 
f(u) = \rho M_U(u) = \frac{\rho}{Z(|U|)} \exp\left\{ \frac{\beta(|U|)}{|U|} \, (u \cdot U) \right\}.  
\label{eq:VMF2}
\end{eqnarray}

Now, with (\ref{eq:VMF2}), the tensor $\Sigma$ can be computed. Introducing again the angle $\theta$, we have: 
\begin{eqnarray*}
\Sigma &=& \int_{u \in {\mathbb S}^1} f \, u \otimes u \, du = \frac{1}{Z} \int_{u \in {\mathbb S}^1} \exp\{ \beta \, (u \cdot \Omega) \} \, (u \otimes u) \, du \\
& = & \frac{\rho}{Z} \int_0^{2 \pi} \left( \begin{array}{cc} \cos^2 \theta & \cos \theta \, \sin \theta \\ \cos \theta \, \sin \theta & \sin^2 \theta \end{array} \right) \, \exp\{ \beta \, \cos \theta \} \, d \theta \\
& = & \frac{\rho}{Z} \int_0^{2 \pi} \left( \begin{array}{cc} \frac{1 + \cos 2 \theta}{2} & 0 \\  0 & \frac{1 - \cos 2 \theta}{2} \end{array} \right) \, \exp\{ \beta \, \cos \theta \} \, d \theta \\
& = & \rho  \left( \begin{array}{cc} \frac{1}{2} (1 + \frac{I_2(\beta)}{I_0(\beta)}) & 0 \\  0 & \frac{1}{2} (1 - \frac{I_2(\beta)}{I_0(\beta)}) \end{array} \right) . 
\end{eqnarray*}
In the third line, the off-diagonal terms have been canceled out because $\sin \theta$ is an odd function of $\theta$. Therefore, introducing
\begin{eqnarray}
&&\hspace{-1cm}
\gamma_\parallel(|U|) = \frac{1}{2 |U|^2} \Big(1 + \frac{I_2(\beta)}{I_0(\beta)} \Big), \qquad  \gamma_\bot(|U|) = \frac{1}{2 |U|^2} \Big(1 - \frac{I_2(\beta)}{I_0(\beta)} \Big), 
\label{eq:gamma}
\end{eqnarray}
with $\beta = \beta(|U|)$, we can write $\Sigma$ as:
\begin{eqnarray}
&&\hspace{-1cm}
\Sigma = \rho \, \big( \, \gamma_\parallel(|U|) \, \, U \otimes U \, + \,  \gamma_\bot(|U|) \, \, U^\bot \otimes U^\bot \, \big) . 
\label{eq:S}
\end{eqnarray}
Since $I_2/I_0 < 1$, the matrix $\Sigma$ is positive definite. The limit $\beta \to 0$ is undefined since $|U| \to 0$ and $\frac{U}{|U|}$ has no definite limit. In the limit $\beta \to \infty$, the matrix converges to $\rho U \otimes U$, and we recover the corresponding term of the monokinetic closure (second term at the left-hand side of (\ref{mk_U})). 

The last term to be computed is the contribution of the force, i.e. the $A_1$ term at the right-hand side of (\ref{eq:A1A2}). Using the first eq. (\ref{eq:mf_rhs}) and (\ref{mf_F}), this contribution is written:
\begin{eqnarray}
&&\hspace{-1cm}
- \int_{u \in {\mathbb S}^1} A_1(x,u,a,t) \, du =  \rho(x,a,t) \, \bar F (x,a,t), \nonumber \\
&&\hspace{-1cm}
\bar F (x,a,t) = \int_{u \in {\mathbb S}^1}   F(x,u,a,t) \,  M_{U(x,a,t)}(u) \, du, 
\label{eq:gauss_F}
\end{eqnarray}
with $F(x,u,a,t)$ given by (\ref{mf_F}). The quantity $\bar F (x,a,t) $, which has a similar physical interpretation as in the monokinetic closure case (see (\ref{mk_F})), is the total force exerted on a ensemble of particles at position $x$ and time $t$ having same target direction $a$. The computation of $F(x,u,a,t)$ follows that of section \ref{subsec:mean_field} and is not repeated here. The only change brought by the VMF closure is in (\ref{mf_D(x)}), which can be written:
\begin{eqnarray}
&& \hspace{-1.5cm} D^{-1}_{(x,u,t)}(w) = \max \, \left\{ \frac{ \int_{y \in S(x,u,t)} \int_{b \in {\mathbb S}^1} E^{-1}(y-x,w,U(y,b,t)) \, \rho(y,b,t)  \,dy \, db}
{ \int_{y \in S(x,u,t)} \int_{b \in {\mathbb S}^1} \rho(y,b,t) \, dy\, db}, \,  \frac{1}{L} \right\}
  . \label{vmf_D(x)}
\end{eqnarray}
with 
\begin{eqnarray}
&& \hspace{-1.5cm} E^{-1}(y-x,w,U) = \int_{v \in {\mathbb S}^1} \tilde D^{-1}(y-x,v-w) \, M_{U}(v) \,dv 
  . \label{vmf_E}
\end{eqnarray}
The quantity $E^{-1}(y-x,w,U)$ can be computed once for all.

We now summarize the macroscopic model based on the VMF closure. It consists of the two equations for the mass and momentum: 
\begin{eqnarray}
& & \hspace{-1cm} 
\partial_t \rho + \nabla_x \cdot (c \rho U) = 0 . 
\label{gauss_rho_2} \\
& & \hspace{-1cm} 
\partial_t (\rho U) + \nabla_x \cdot \big( \,  c \rho  \, \big( \, \gamma_\parallel \, U \otimes U \, + \,  \gamma_\bot \, U^\bot \otimes U^\bot \, \big) \big) =  \rho \, \bar F - d \rho U, \label{gauss_U_2}  
\end{eqnarray}
together with the expression (\ref{eq:gauss_F}) of the force $\bar F$. We have omitted the dependences of the coefficients $\gamma$ upon $|U|$. This is a system for $\rho(x,a,t)$ and $U(x,a,t)$. The target direction $a$ appears implicitly through the expression of the force $\bar F$ which couples all target directions together. The properties of this system, and particularly, its hyperbolicity, will be studied in future work.

\subsubsection{VMF closure: discussion}
\label{subsubsec_gaussian_discussion}

The interpretation of the two equations (\ref{gauss_rho_2}), (\ref{gauss_U_2}) is the same as for the monokinetic closure (see discussion at section \ref{subsubsec_monokinetic_discussion}). Eq. (\ref{gauss_rho_2}) expresses the mass conservation, while eq. (\ref{gauss_U_2}) describes how the mean velocity evolves in time. Compared to the monokinetic closure, Eq. (\ref{gauss_U_2}) is more naturally written in terms of the momentum $\rho U$ and presents three major differences. 

The first one is the expression of the transport term $\nabla_x \cdot \big( c \rho  \, (\gamma_\parallel \, U \otimes U \, + \,  \gamma_\bot \, U^\bot \otimes U^\bot ) \big)$, which, compared to (\ref{mk_U}),  involves two terms. The first one, proportional to $U \otimes U$ is similar to the term involved in (\ref{mk_U}), but is multiplied by a coefficient $\gamma_\parallel (|U|)$ which is less than $1$. The second term, proportional to the tensor $U^\bot \otimes U^\bot$, is unusual in fluid models. It arises from the VMF closure, itself justified by the fact that microscopic velocities are constrained to be of norm $1$. For such fluids, significant differences from classical fluid dynamics models have already been found \cite{Degond_etal_arXiv:1109.2404, Degond_Motsch_M3AS08}. Similar unconventional models have been found even if the microscopic velocities are not constrained to be of norm $1$ when the particles are subject to a self-propulsion force \cite{Barbaro_Degond_arXiv:1207.1926}. The impact of this new term on the hyperbolicity of the model will be studied in future work. 

The second difference is the presence of a velocity damping term (second term at the right-hand side of (\ref{gauss_U_2})). This damping term is produced by the noise. At the kinetic level, the dynamics induced by the velocity diffusion operator is that of the heat equation on the circle ${\mathbb S}^1$. It makes the velocity distribution function more and more isotropic. Under this dynamics, the average velocity tends to zero which is what the damping term at the right-hand side of (\ref{gauss_U_2}) expresses. 

The third difference is in the computation of the force term $\bar F$. It is constructed by averaging elementary force terms over the probability distribution $M_U(u)$ (see (\ref{eq:gauss_F})). The potential of each elementary force term depends on the averaged DTI of the particles belonging to the corresponding fluid element. This averaged DTI involves the average of the elementary DTI over the probability distribution $M_U(u)$ again (formulas (\ref{vmf_D(x)}) and (\ref{vmf_E})). The result of this procedure is a non-local expression of the force involving both the mass and velocity distributions $\rho$ and $U$. 

We can derive a local version of the interaction force, using the local kinetic framework of section \ref{subsec:mean_field_local}. In this framework, the averaged DTI is given by:
\begin{eqnarray}
&& \hspace{-1cm} D^{-1}_{(x,u,t)}(w) = \max \, \left\{ \frac{ \int_{b \in {\mathbb S}^1} {\mathcal E}_{\kappa, \delta(x,t)}^{-1}(u,w,U(x,b,t))  \, \rho(x,b,t) \, db}
{ \int_{b \in {\mathbb S}^1} \rho(x,b,t)  \, db} , \,  \frac{1}{L} \right\}
  , \label{vmf_D(x)_local}
\end{eqnarray}
with
\begin{eqnarray}
&& \hspace{-1cm} {\mathcal E}_{\kappa, \delta}^{-1}(u,w,U) =  \int_{v \in {\mathbb S}^1} \Delta^{-1}_{\kappa, \delta} (u,v-w) \, M_{U}(v) \,  dv   . 
\label{vmf_E(x)_local}
\end{eqnarray}
The function ${\mathcal E}_{\kappa, \delta}$ can be computed numerically. 

In the special case $\kappa = -1$ (no restriction of the interaction region to a forward vision cone), the function $\Delta_{\kappa, \delta}$ becomes independent of $u$ and the formulas simplify into
\begin{eqnarray}
&& \hspace{-1cm} D^{-1}_{(x,t)}(w) = \max \, \left\{ \frac{ \int_{b \in {\mathbb S}^1} {\mathcal E}_{\delta(x,t)}^{-1}(w,U(x,b,t))  \, \rho(x,b,t) \, db}
{ \int_{b \in {\mathbb S}^1} \rho(x,b,t)  \, db} , \,  \frac{1}{L} \right\}
  , \label{vmf_D(x)_local_2}
\end{eqnarray}
with
\begin{eqnarray}
&& \hspace{-1cm} {\mathcal E}_{\delta}^{-1}(w,U) =  \int_{v \in {\mathbb S}^1} \Delta^{-1}_{\delta} (|v-w|) \, M_{U}(v) \,  dv   , 
\label{vmf_E(x)_local_2}
\end{eqnarray}
and $\Delta_{\delta} (|v-w|)$ given by (\ref{mf_Delta_local}). Again, the function ${\mathcal E}_{\delta}^{-1}(w,U)$ can be numerically computed once for all. We note that $D^{-1}_{(x,t)}(w)$ does not explicitly depend on $u$. It results that the potential does not depend explicitly on $u$ either, and can be written 
\begin{eqnarray}
&& \hspace{-1cm} \Phi_{(x,a,t)}(w) = \frac{k}{2} |D_{(x,t)}(w) w - L \, a|^2  . \label{vmf_Phi_local_2}
\end{eqnarray}
The elementary force can now be expressed by a gradient with respect to the $u$ variable: 
\begin{eqnarray}
&& \hspace{-1cm} F(x,u,a,t) = - \nabla_u \Phi_{(x,a,t)}(u)  . \label{vmf_F_local_2}
\end{eqnarray}
This was not possible before. Indeed, because of the explicit dependence of $\Phi$ on $u$, we had to distinguish between the variables $u$ and $w$ in the expression of the force (\ref{mf_F}). In the present case, thanks to (\ref{vmf_F_local_2}), eq. (\ref{eq:gauss_F}) for the force $\bar F (x,a,t)$ can be simplified. Indeed, using Stokes formula in (\ref{eq:gauss_F}) together with (\ref{vmf_F_local_2}), we get, for each component $k=1,2$ and denoting by $e_k$ the $k$-th basis vector: 
\begin{eqnarray*}
\bar F_k (x,a,t) &=& - \left( \int_{u \in {\mathbb S}^1}   \nabla_u \Phi_{(x,a,t)}(u) \,  M_{U(x,a,t)}(u) \, du \right) e_k\\
&=& - \int_{u \in {\mathbb S}^1}   \nabla_u \Phi_{(x,a,t)}(u) \,  M_{U(x,a,t)}(u) \, (e_k \cdot u^\bot) u^\bot \, du  \\
&=&  \int_{u \in {\mathbb S}^1}   \Phi_{(x,a,t)}(u) \, \nabla_u \cdot  \big( M_{U(x,a,t)}(u) \, u_k^\bot \, u^\bot \big) \, du, 
\end{eqnarray*}
which, after easy computations, leads to
\begin{eqnarray}
&& \hspace{-1cm}
\bar F (x,a,t) = \beta \big(|U(x,a,t)| \big)  \int_{u \in {\mathbb S}^1}  \Phi_{(x,a,t)}(u)  \,  M_{U(x,a,t)}(u) \, du \, \frac{U(x,a,t)}{|U(x,a,t)|}  \nonumber \\
&& \hspace{-0.6cm} 
-   \int_{u \in {\mathbb S}^1}  \Phi_{(x,a,t)}(u)  \,  M_{U(x,a,t)}(u) \, \Big(1 + \beta \big(|U(x,a,t)| \big) \big( u \cdot \frac{U(x,a,t)}{|U(x,a,t)|}\big)\Big) \, u \, du . 
\label{eq:gauss_F_local_2}
\end{eqnarray}
Inserting (\ref{vmf_Phi_local_2}) into (\ref{eq:gauss_F_local_2}) leads to an alternate expression of the force (in the case $\kappa=-1$).

\subsection{Hydrodynamic limit}
\label{subsec:hydro}

\subsubsection{Hydrodynamic limit: derivation}
\label{subsubsec_hydro_derivation}

Here, we consider the situation of section \ref{subsec:mean_field_local}. We assume that the various length scales associated to the interactions between pedestrians are very small. Additionally, we assume that $\kappa = -1$, i.e. the interaction region is the disk $B_\delta$ and there is no blind zone behind the subjects (see (\ref{eq:Bdelta})). In this case (see bottom of section \ref{subsec:mean_field_local}), the potential does not explicitly depend on $u$. In the present section, we take advantage of this simplification to perform the hydrodynamic limit of the mean-field model (\ref{mf_f}). The hydrodynamic limit consists in supposing that the changes in pedestrian velocities due either to the interaction force $F$ or to the noise with diffusion constant $d$ occur on very short time scales. To highlight this scaling assumption, we introduce a small parameter $\varepsilon \ll 1$ and rescale the force and diffusion constants in such a way that 
\begin{equation} 
F = \frac{1}{\varepsilon} \hat F, \qquad d = \frac{1}{\varepsilon} \hat d. 
\label{eq:scaling_hydro}
\end{equation}

Dropping the 'hats' for simplicity and writing all unknowns with a superscript $\varepsilon$, we can write the mean-field model (\ref{mf_f}) with local interaction force and no blind zone as follows: 
\begin{eqnarray}
&& \hspace{-1cm} 
\partial_t f^\varepsilon + c u \cdot \nabla_x f^\varepsilon = \frac{1}{\varepsilon} Q_{D_{f^\varepsilon}}(f^\varepsilon), 
\label{eq:hydro_f}
\end{eqnarray}
where the operator $Q_{D_f}(f)$ (the  so-called 'collision operator' of kinetic theory) describes the rate of change of the pedestrian velocities due to their interaction with the other pedestrians and the noise. We parametrize the collision operator by the DTI $D_f$.  Because of the local interaction assumption, $Q_{D_f}(f)$ operates only on $u$ and $a$, leaving $(x,t)$ untouched. Therefore, we describe it as an  operator acting on functions of $f(u,a)$ only. For a given function $u \in {\mathbb S}^1 \to D(u) \in {\mathbb R}_+$, the collision operator is written
\begin{eqnarray}
&& \hspace{-1cm} 
Q_D(f) = -  \nabla_u \cdot ( F_D \, f) + d \Delta_u f , 
\label{eq:hydro_Q}
\end{eqnarray}
where the force $F_D(u,a)$ is written in terms of the potential $\Phi_D(u,a)$ by: 
\begin{eqnarray}
&& \hspace{-1cm} 
F_D (u,a) = -  \nabla_u \Phi_D(u,a) , \quad  \Phi_D(u,a) = \frac{k}{2} |D(u) u - L \, a|^2  . 
\label{eq:hydro_F_Phi}
\end{eqnarray}
For a given function $f(u,a)$, $D_f(u)$ is defined by 
\begin{eqnarray}
&& \hspace{-1cm} D^{-1}_f(u) = \max \, \left\{ \frac{ \int_{(v,b) \in {\mathbb T}^2} \Delta^{-1}_{\delta_f} (|v-u|) \, f(v,b) \, dv \, db}
{ \int_{(v,b) \in {\mathbb T}^2} f(v,b) \, dv \, db} , \,  \frac{1}{L} \right\}
  , \label{eq:hydro_D(x)_local}
\end{eqnarray}
and $\Delta_{\delta} (|v-u|)$ is the known function given by (\ref{mf_Delta_local_iso}). Finally, $\delta_f$ is given by
\begin{eqnarray}
&& \hspace{-1cm} 
\delta_f = C \, \Big( \int_{(v,b) \in {\mathbb T}^2} f(v,b) \, dv \, db \Big)^{-1/2}. 
\label{eq:delta}
\end{eqnarray}
It is an immediate matter to check that this sequence of definitions is equivalent to the mean-field model (\ref{mf_f}) with local interaction force (\ref{mf_F}), (\ref{mf_Phi}) and (\ref{mf_D(x)_local}), in the case $\kappa = -1$ (up to the change of scale (\ref{eq:scaling_hydro})). Here, we note that the first equation (\ref{eq:hydro_F_Phi}) is equivalent to (\ref{mf_F}), because in the case $\kappa = -1$, the potential $\Phi_{x,a,t}(w)$ does not explicitly depend on $u$. These two formulas are not equivalent if the potential  (\ref{mf_Phi}) depends explicitly on $u$, which is the case when $\kappa >-1$. This is why this section is restricted to the case $\kappa =-1$.

Now, by inspecting (\ref{eq:hydro_f}) in the limit $\varepsilon \to 0$, we deduce that $f^\varepsilon \to f^0$ where $f^0$ is  the solution of 
\begin{eqnarray}
&& \hspace{-1cm} 
Q_{D_{f^0}} (f^0) = 0 . 
\label{eq:null_space_Q}
\end{eqnarray}
Any such solution is called a Local Thermodynamical Equilibrium (LTE).  By the fact that the operator $f \to Q_{D_{f}} (f)$ only operates on $(u,a)$, this equation specifies the dependence of $f^0$ on $(u,a)$, leaving the dependence on $(x,t)$ undetermined at this level. 

\begin{figure}[htbp]
\begin{center}
\input{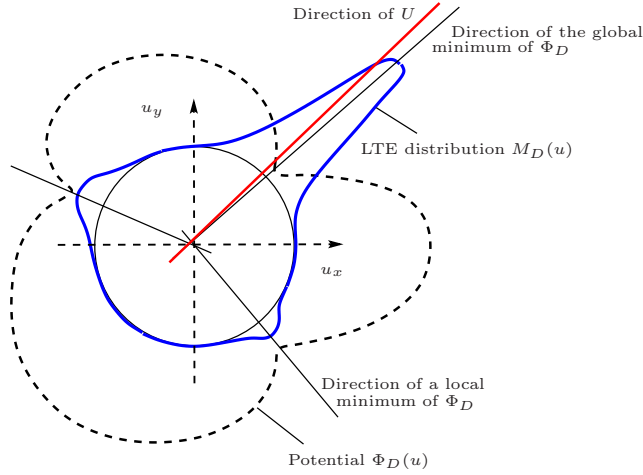}
\caption{The LTE distribution $u \in {\mathbb S}^1 \to M_D(u,a)$ for a given $a \in {\mathbb S}^1$ as a function of $u$ in polar coordinates (blue curve). The distribution $M_D$ is maximal where the potential $\Phi_D$ (black dashed curve) is minimal. The minima of $\Phi_D$ and maxima of $M_D$ are indicated by black semi-lines. The global maximum of $M_D$ corresponds to the global minimum of $\Phi_D$. The distribution $M_D$ has bell-like shapes around its maxima. Their width are roughly proportional to the noise level $\sqrt d$. The direction of the mean velocity $U$ is given by the red semi-line. It is fully determined by $M_D$ and therefore, by $\Phi_D$ and is a function of $(x,a,t)$ like $\Phi_D$. We have $|U| < 1$. }
\label{fig:Nash}
\end{center}
\end{figure}

In order to determine the LTE's, we first suppose that $D$: $u \in {\mathbb S}^1 \to D(u) \in {\mathbb R}_+$ is a given function. We note that for a fixed function $D$, the operator $f \to Q_D(f)$ is a linear Fokker-Planck operator. We introduce the function 
\begin{eqnarray}
&& \hspace{-1cm} 
M_D(u,a) = \frac{1}{Z_D(a)} \exp \big( - \frac{ \Phi_D(u,a) }{d} \big) ,  
\label{eq:hydro_LED}
\end{eqnarray}
with $Z_D(a)$ the normalizing constant such that 
\begin{eqnarray}
&& \hspace{-1cm} 
\int_{u \in {\mathbb S}^1} M_D(u,a) \, du = 1. 
\label{eq:hydro_LED_normal}
\end{eqnarray}
The function $u \in {\mathbb S}^1 \to M_D(u,a)$ for a given $a \in {\mathbb S}^1$ is represented graphically on Fig.~\ref{fig:Nash} (blue curve) as a function of $u$ in polar coordinates. We notice that $M_D$ is maximal where the potential $\Phi_D$ (represented by the black dashed curve) is minimal. Around its maxima, the graphical representation of $M_D$ is bell-shaped. The corresponding widths are roughly proportional to the noise level $\sqrt d$.

Thanks to (\ref{eq:hydro_F_Phi}), we can write:
\begin{eqnarray}
&& \hspace{-1cm} 
Q_D(f) = -  d \, \nabla_u \cdot \Big( M_D \nabla_u \big( \frac{f}{M_D} \big) \Big). 
\label{eq:hydro_Q_2}
\end{eqnarray}
By applying Green's formula, we deduce that for any function $f(u,a)$ with appropriate regularity, we have:
\begin{eqnarray}
&& \hspace{-1cm} 
\int_{(u,a) \in {\mathbb T}^2} Q_D(f) \, \frac{f}{M_D} \, du \, da =  
- \int_{(u,a) \in {\mathbb T}^2} M_D \, \Big|  \nabla_u \big( \frac{f}{M_D} \big) \Big|^2 \, du \, da.  
\label{eq:hydro_intQ}
\end{eqnarray}
Therefore, the equation 
\begin{eqnarray}
&& \hspace{-1cm} 
Q_D(f) = 0 , 
\label{eq:hydro_nullQ_1}
\end{eqnarray}
is equivalent to saying that there exists a function $\rho$: $a  \in {\mathbb S}^1 \to \rho(a) \geq 0$ such that 
\begin{eqnarray}
f(u,a) = \rho(a) \, M_D(u,a). 
\label{eq:hydro_nullQ_2}
\end{eqnarray}
Indeed, inserting (\ref{eq:hydro_nullQ_2}) into (\ref{eq:hydro_Q_2}) clearly leads to (\ref{eq:hydro_nullQ_1}). Reciprocally, if (\ref{eq:hydro_nullQ_1}) is satisfied, then, applying (\ref{eq:hydro_intQ}) leads to the fact that $\nabla_u ( f/M_D) =0$, i.e. $f/M_D$ is a function of $a$ only, which is exactly saying (\ref{eq:hydro_nullQ_2}). 

Therefore, an LTE is necessarily of the form (\ref{eq:hydro_nullQ_2}). However, there is a consistency condition to be satisfied. For $f = \rho M_D$ to be a solution of (\ref{eq:null_space_Q}), we need to ensure that $D=D_f$. In other words, $D$ is not arbitrary, but must be the DTI associated to $f$. Inserting (\ref{eq:hydro_nullQ_2}) into (\ref{eq:hydro_D(x)_local}), this constraint is written: 
\begin{eqnarray}
&& \hspace{-1cm} D^{-1}(u) = \max \, \left\{ \frac{ \int_{(v,b) \in {\mathbb T}^2} \Delta^{-1}_{\delta_{\rho M_D}} (|v-u|) \, \rho(b) \, M_D(v,b) \, dv \, db}
{ \int_{b \in {\mathbb S}^1} \rho(b) \, db} , \,  \frac{1}{L} \right\}
  . \label{eq:hydro_Drho}
\end{eqnarray}
For any function $\rho$: $a  \in {\mathbb S}^1 \to \rho(a) \geq 0$ we look for functions $D_\rho$: $u  \in {\mathbb S}^1 \to D_\rho(u) \geq 0$ of this functional equation. The questions whether it admits a solution and how many such solutions exist is left to future work. Here,  we assume that there exists at least one isolated branch of solutions $D_\rho$. Therefore, the LTE are of the form $\rho \, M_{D_\rho}$, with $D_\rho$ a solution of (\ref{eq:hydro_Drho}). By restoring the dependence upon $(x,t)$, we conclude that the solutions of (\ref{eq:null_space_Q}) are of the form 
\begin{eqnarray}
f^0(x,u,a,t) = \rho_{(x,t)}(a) \, M_{D_{\rho_{(x,t)}}}(u,a), 
\label{eq:hydro_LED_2}
\end{eqnarray}
where, for any $(x,t)$, the function $u \to D_{\rho_{(x,t)}}(u)$ satisfies (\ref{eq:hydro_Drho}) with $\rho(b)$ replaced by $\rho_{(x,t)}(b)$. By the normalization condition (\ref{eq:hydro_LED_normal}), $\rho_{(x,t)}(a)$ appears as the density of pedestrians at point $x$ and time $t$ with target velocity $a$. It has the same meaning as $\rho(x,t,a)$ in the monokinetic or VMF closures, but is written differently to highlight its dependence on the target velocity $a$. Indeed, the LTE at point $(x,t)$ depends functionally on the function $a \to \rho_{(x,t)}(a)$. By contrast, the LTE at $(x,t)$ does not depend on the function $(x,t) \to \rho_{(x,t)}(a)$. This motivates this dissymetric treatment of the dependences of $\rho$ on $a$ on the one hand and $(x,t)$ on the other hand. Now, we are looking for the equations specifying how $\rho_{(x,t)}(a)$ varies with $(x,t)$.

To do so, we first notice that, because of the divergence form of (\ref{eq:hydro_Q_2}), we have 
\begin{eqnarray}
&& \hspace{-1cm} 
\int_{u \in {\mathbb S}^1} Q_{D_f}(f) \, du = 0. 
\label{eq:hydro_intQ=0}
\end{eqnarray}
Therefore, integrating (\ref{eq:hydro_f}) with respect to $u$ and using (\ref{eq:hydro_intQ=0}) leads to 
\begin{eqnarray}
&& \hspace{-1cm} 
\partial_t \rho^\varepsilon + \nabla_x \cdot (c \rho^\varepsilon U^\varepsilon) = 0, 
\label{eq:hydro_rho}
\end{eqnarray}
with $\rho^\varepsilon(x,a,t)$ the density and $U^\varepsilon(x,a,t)$ the mean velocity of pedestrians at position $x$, time $t$ and target direction $a$, given by:
\begin{eqnarray}
&& \hspace{-1cm} 
\rho^\varepsilon(x,a,t) =  \int_{u \in {\mathbb S}^1} f^\varepsilon(x,u,a,t) \, du , \quad (\rho^\varepsilon U^\varepsilon)(x,a,t) = \int_{u \in {\mathbb S}^1} f^\varepsilon(x,u,a,t) \, u\, du. 
\label{eq:hydro_mom}
\end{eqnarray}
Eq. (\ref{eq:hydro_rho}) is the continuity equation for pedestrians of target velocity $a$ and is valid all the time (i.e. even when $\varepsilon$ is not small). 
Now, taking the limit $\varepsilon \to 0$ in (\ref{eq:hydro_mom}) and using the fact that $f^\varepsilon \to f^0$, where $f^0$ satisfies (\ref{eq:hydro_LED_2}), we get
\begin{eqnarray}
&& \hspace{-1cm} 
\rho^\varepsilon(x,a,t)  \to \rho_{(x,t)}(a), \quad U^\varepsilon(x,a,t) \to U_{\rho_{(x,t)}}(a), 
\label{eq:hydro_mom_lim}
\end{eqnarray}
with 
\begin{eqnarray}
&& \hspace{-1cm} 
U_{\rho}(a) = \int_{u \in {\mathbb S}^1} M_{D_{\rho}}(u,a) \, u\, du. 
\label{eq:hydro_vel}
\end{eqnarray}
Of course, taking the limit $\varepsilon \to 0$ in the continuity eq. (\ref{eq:hydro_rho}) leads to 
\begin{eqnarray}
&& \hspace{-1cm} 
\partial_t \rho_{(x,t)}(a) + \nabla_x \cdot (c \rho_{(x,t)}(a) U_{\rho_{(x,t)}}(a)) = 0.
\label{eq:hydro_rho_lim}
\end{eqnarray}

To summarize, the hydrodynamic model provides the spatio-temporal evolution of the density $\rho_{(x,t)}(a)$ of pedestrians with target direction $a$. It consists of the single continuity eq. (\ref{eq:hydro_rho_lim}). The different target directions are coupled together through the computation of the average velocity $U_{\rho}(a)$ by means of (\ref{eq:hydro_vel}). It requires the determination of the DTI at the same point. The DTI $D_\rho(u)$ is the solution of the functional equation (\ref{eq:hydro_Drho}), parametrized by the function $\rho(a)$. This computation is local in space-time but must be realized at any discretization point in space-time $(x,t)$. Therefore, the practical determination of the velocity $U_{\rho_{(x,t)}}(a)$ may require high computational power. However, the local character of the problem is perfectly adapted to massively parallel or graphical computers.  Note that this hydrodynamic model belongs to the class of first order models of traffic, since the velocity is fully determined by the knowledge of the density.

\subsubsection{Hydrodynamic limit: discussion}
\label{subsubsec_hydro_discussion}

The rationale of this model is best understood if a time discretization is performed. Suppose that the distribution $\rho^n_x(a)$ of pedestrians at position $x$ and target direction $a$ is known at time $t^n = n \Delta t$. We update this density at time $t^{n+1}$ using the following time-discrete version of the continuity eq. (\ref{eq:hydro_rho_lim}) by the Euler method:
\begin{eqnarray}
&& \hspace{-1cm} 
\rho^{n+1}_x(a) = \rho^n_x(a) - \Delta t  \, \nabla_x \cdot (c \rho^n_x(a) U_{\rho^n_x}(a)) = 0.
\label{eq:hydro_rho_lim_disc}
\end{eqnarray}
To use this scheme, it is necessary to compute the velocity $U_{\rho^n_x}(a)$. For this purpose, the DTI $D_{\rho^n}(u)$ need to be computed by solving the functional equation (\ref{eq:hydro_Drho}), where $\rho^n$ is substituted for $\rho$. Once $D_{\rho^n}$ is known, the LTE (\ref{eq:hydro_LED_2}) can be computed and as a by-product, the mean velocity $U_{\rho^n_{(x)}}(a)$ through (\ref{eq:hydro_vel}) (see Fig. \ref{fig:Nash}: the mean velocity is represented by the red semi-line). 

The heart of the model is the process of finding the velocities, given the density $\rho^n$ of pedestrians having prescribed target velocities. This process is decomposed as follows. First, knowing the density $\rho^n$, the DTI in all directions are computed by solving the functional equation (\ref{eq:hydro_Drho}). We note that the DTI $D_{\rho^n}(u)$ is independent of the target direction $a$ and only describes the ability of a pedestrian to move in the direction $u$, given the density $\rho^n$. The functional equation (\ref{eq:hydro_Drho}) describes how each pedestrian optimizes his actual velocity, i.e.  minimizes the potential $\Phi_{D_\rho}$, taking into account all other pedestrians around. This functional equation expresses a Nash equilibrium of a  non-cooperative anonymous game with a continuum of players. Such games are characterized by an infinite number of players forming a continuum \cite{Aumann_Econometrica64}. They are non-cooperative i.e. they exclude the possibility for the players to cooperate to improve their gain \cite{Green_Porter_Econometrica84}. Finally, they are anonymous in the sense that two players with the same strategy cannot be distinguished \cite{Schmeidler_JStatPhys73}. Recently, this category of games has been at the heart of the theory of 'Mean-Field Games' \cite{Lasry_Lions_JapanJMath07}. Here, the strategy variable of the players is the velocity $u$, while the target direction $a$ is the players' type (see an introduction to game theory in \cite{Blanchet_HDR12}). The players' utility function is the opposite of the potential $\phi_D(u,a)$. The functional eq. (\ref{eq:hydro_Drho}) expresses that each pedestrian separately cannot improve his utility function by choosing a different velocity $u$, which is the definition of a Nash equilibrium. This model is a particular example of the framework relating game theory and kinetic theory developed in \cite{Degond_etal_arXiv:1212.6130}. This viewpoint will be further expanded in future work. 

We stress the local character of the model: this optimum is realized locally, i.e. at any point $x$ and at all times $t$. Once the equilibrium $D_{\rho^n}$ has been found, the LTE and the mean velocity $U_{\rho^n}$ follow directly. We note that the dependence of $U_\rho$ upon $\rho$ is functional, i.e. the value $U_\rho(a)$ for a given target velocity $a$ depends on $\rho(b)$ for all values of $b \in {\mathbb S}^1$. This can be understood easily. If there are more pedestrians heading towards a given direction, say $b_0$, the DTI will be affected in all directions $u$ and correlatively, the mean-value of the velocity of pedestrians heading towards direction $a$ will be changed, even if $a$ is very different from $b_0$.

\setcounter{equation}{0}
\section{Comparison with previous work}
\label{sec:discussion}

In this section, we compare our results to the literature. We refer to \cite{Moussaid_etal_PLOSCB12, Moussaid_etal_PNAS11} for a discussion of the original discrete IBM. The time-continuous IBM (section \ref{subsec:IBM_time_continuous}) obviously bears analogies with the social force model \cite{Helbing_BehavSci91, Helbing_Molnar_PRE95, Helbing_Molnar_SelfOrganization97}. However, in our model, the elementary binary interactions are combined in a non-linear way (i.e. they are  nonlinearly additive, see \cite{Bellomo_Dogbe_SIAMRev11, Moussaid_etal_PNAS11} for a discussion of this point). The velocity potential (\ref{Phi_i}) is reminiscent of the 'steering potential' model of \cite{Huang_etal_RoboticsAutonomSyst06}. However, in \cite{Huang_etal_RoboticsAutonomSyst06}, the potentials of the various obstacles are added linearly by contrast to the present work, as already mentioned. Analogies also exist with the optimal control model of \cite{Hoogendoorn_Bovy_OptControlApplMeth03}. Indeed, in our work, the potential is similar to a cost function that the pedestrian dynamic tends to minimize. In  \cite{Hoogendoorn_Bovy_OptControlApplMeth03}, three types of costs are considered: (i) the cost of drifting away from the planned trajectory, (ii) the cost of walking too close to other pedestrians and (iii) the cost of acceleration. In our constant velocity model, we have not considered any cost associated to accelerating (i.e. turning), which is probably incorrect. We also replace cost (ii) by a constraint (the distance traveled in direction $w$ cannot exceed the DTI $D_i(w)$), and we minimize cost (i) subject to this constraint. The smoother expression of cost (ii) in \cite{Hoogendoorn_Bovy_OptControlApplMeth03} allows for the inclusion of advanced features, such as the body compressibility. However, adding more features increases the number of parameters that need to be calibrated from the data. Our model has quite few parameters which need to be calibrated, which is an advantage in the context of scarce noisy data. 

We now turn to the mean-field kinetic model (section \ref{subsec:mean_field}). As already mentioned, kinetic models are scarce in the literature \cite{Bellomo_Dogbe_SIAMRev11}. Ref. \cite{Bellomo_Bellouquid_MathModelCollectivBehav10} proposes a general kinetic framework for traffic and crowd dynamics but the specific features of pedestrian interactions are not detailed. The model presented in section \ref{subsec:mean_field} seems to be one of the very first crowd kinetic models based on a detailed analysis of pedestrian behavior. 

Most fluid models for crowds have been envisioned as extensions of road traffic models. Fluid models for traffic roughly fall in two categories: (i) first-order models which are composed of the continuity equation and an algebraic equation relating the flux to the density and (ii) second-order models, where the continuity equation is complemented with an evolution equation for the mean velocity. The prototype of first-order models is the Lighthill-Whitham-Richards (LWR) model \cite{Lighthill_Whitham_ProcRoySocA55}, while second-order models are represented by the Payne-Whitham (PW) \cite{Payne_SimCoun71} and Aw-Rascle (AR) \cite{Aw_Rascle_SIAP00} models. Clearly, our first two fluid models, the monokinetic closure (section \ref{subsec_monokinetic}) and the VMF closure (section \ref{subsec:gaussian}) belong to the class of second-order models, while the third one, obtained through the hydrodynamic limit (section \ref{subsec:hydro}), is a first order model. 

We first discuss the hydrodynamic model of section \ref{subsec:hydro} in reference to the literature on first-order models. The difficulty with extending the LWR model of traffic to crowds is the passage from one to two dimensions in the prescription of the flux. While the traffic flux is a scalar quantity, the pedestrian flux is a vector and a prescription is needed to decide of its direction. In this respect, there are three classes of first order crowd models: (i) those where the direction of the flux is fixed locally as a function of the density or its gradients \cite{Berres_etal_NHM11, Coscia_Canavesio_M3AS08}, (ii) those where the direction is fixed by a non-local average \cite{Piccoli_Tosin_ContMechThermo09} and (iii) those where the direction is given through the solution of a Partial Differential Equation (PDE), such as the Eikonal equation \cite{Huang_etal_TranspResB09, Hughes_TranspResB02, Hughes_AnnRevFluidMech03}. Our model bears the strongest analogies with the third type. Indeed, in the determination of the flow direction the Eikonal equation is replaced by the functional equation (\ref{eq:hydro_D(x)_local}), which offers a similar level of implicitness. Additionally, in \cite{Hughes_TranspResB02}, the pedestrians minimize their travel times through an estimate which gives more weight to crowded areas. Therefore, the optimization principles underlying the dynamics of both our model and \cite{Hughes_TranspResB02} are similar. However, in \cite{Hughes_TranspResB02}, only the total density is taken into account in the travel time estimation, while our model also includes information about the velocities and target velocities. In particular, the following behavior (i.e. the fact that a pedestrian does not necessarily see a pedestrian moving in the same direction as himself as an obstacle) is likely to be better taken into account in our model. 

Second-order models for crowds are more scarce. Basically, referring to the classification in the paragraph above, only classes (i) (local prescription of the direction of the flux within the PW model \cite{AlNasur_Kachroo_IEEEITSC06} or AR model \cite{Bellomo_Dogbe_M3AS08}) and (iii) (coupling of a PW model with an eikonal equation for the flux direction \cite{Jiang_etal_PhysicaA10}) have been explored. The models of \cite{Helbing_ComplexSyst92, Henderson_TranspRes74} may be seen as belonging to class (i), although they involve a third equation (the energy balance equation). Our monokinetic and VMF closure models (sections \ref{subsec_monokinetic} and \ref{subsec:gaussian}) bear the strongest analogies with these last two references as they are obtained using similar methodologies (by closing a moment hierarchy from a kinetic equation). The least difference arises in the case of the monokinetic closure, which is close to a zero-temperature fluid equation. The non-local expression of the force or its local approximation still make the specificity of our model. In \cite{Helbing_ComplexSyst92, Henderson_TranspRes74}, the effects of the interactions between the pedestrians are mostly embedded in the energy balance equation. The VMF closure offers more differences with the fluid mechanical view of \cite{Helbing_ComplexSyst92, Henderson_TranspRes74}. Indeed, the latter postulate a Maxwellian (i.e. Gaussian) velocity distribution of pedestrians. This is obviously unlikely since pedestrians cannot reach arbitrarily large velocities. Our microscopic dynamics, which constrains the velocities to be of constant norm, more closely mimics the real behavior of a pedestrian (at least at moderate densities where he is able to walk most of the time at his free speed). As a consequence of this velocity constraint, the closure cannot be Maxwellian, but is a VMF distribution instead. We have seen in section \ref{subsubsec_gaussian_discussion} that very specific features emerge from this unusual 'hydrodynamics'. Obviously, these features are not taken into account in standard fluid models. 

Finally, the last point of our discussion is a comparison between the three models presented in this paper. The third one, which is related to game-theoretic concepts, captures nicely the mechanisms by which pedestrians achieve a consensus and maintain traffic efficiency even in very crowded environments. However, it is restricted to local interactions and uniform vision (i.e. no blind zone). Although actual pedestrians do have a blind zone, the approximation involved in the "no-blind zone" model may find its justification as follows:  pedestrians walking behind the subject are unlikely to significantly modify the value of his DTI. Indeed the threat of a collision of the subject with these pedestrians is weak, since both are walking with the same speed. Therefore, the DTI is identical, weather its computation includes all pedestrians in the neighbourhood of the subject or excludes those who are inside the blind zone. Consequently, there is little approximation involved in replacing the actual vision cone by the entire neighbourhood of the subject. Considering interactions within their local approximation makes the model only suitable to the large-scales, where the interaction region of the subjects is small compared to the size of the scene. Another drawback of the model is the complexity of solving the fixed point equation for the DTI everywhere in space-time. Unless a fast algorithm is found and massively computers are used, this can lead to overwhelming computer costs. 

For this reason the second model, which relies on the VMF closure constitutes a good compromise between physical accuracy and computational efficiency. Its usability however is subjected to its well-posedness, i.e. to its hyperbolicity, a property which still remains to be investigated. Finally, the first model, which relies on the monokinetic closure, is the simplest one. At least, it guarantees local-in-time well-posedness, i.e. the model has a unique solution until two pedestrian trajectories issued from initially different points meet. At such a meeting point, a mass concentration occurs, which is obviously unphysical. A way out this unpleasant feature would be to restore non constant speeds and to allow pedestrians to slow down in case of a close encounter, which they obviously do in practice. Consideration of non-constant speeds is the subject of future work.

\setcounter{equation}{0}
\section{Conclusion}
\label{sec:conclu}

In this article, we propose a hierarchy of macroscopic models derived from the heuristic behavioral Individual-Based Model of \cite{Moussaid_etal_PNAS11} and discuss them in view of the available literature. This IBM supposes that pedestrians optimize their trajectory towards their goal subject to the constraint of avoiding collisions with neighboring pedestrians. We first propose a novel kinetic model. In a second step, we derive three different fluid models. The first two ones consist of balance equations for the density and mean velocity of the pedestrians. They differ by the proposed closure relations, based on a monokinetic Ansatz for the first one and on the VMF distribution for the second one. The third model, which has more restrictive assumptions, consists of a single mass conservation equation, with mass flux functionally related to the density distribution. The functional relation expresses the realization of a Nash equilibrium where each pedestrian finds his optimal direction of motion towards his target in the midst of the other pedestrians and in the presence of noise. These models are the first available kinetic and fluid models derived from the heuristic behavioral Individual-Based Model of \cite{Moussaid_etal_PNAS11}. Future work will be devoted to the study of these various models, both from the theoretical and numerical viewpoints, and to their confrontation with the experimental data.


\end{document}